# Modelling of Polymer Clay Nanocomposites for a Multiscale Approach

P. E. SPENCER and J. SWEENEY

*IRC in Polymer Science & Technology, School of Engineering Design and Technology, University of Bradford, Bradford, UK*

## 1. INTRODUCTION

Polymer nanocomposites have been in existence for over 20 years. Recently, Okada & Usuku [1] have presented a historical review and Hussain et al. [2] have reviewed the current science and technology. Filler particles, with at least one dimension at the nanometre level, are embedded within a polymer matrix, and can effect significant improvements in mechanical properties – modulus, yield stress, fracture toughness – when present at quite low levels of a few percent by weight (wt. %). This is associated with particle geometries of very high aspect ratio and resultant high surface areas per unit volume. Significant improvements are obtained at filler concentrations so low that the optical properties of the polymer matrix are largely unaffected, so that a transparent matrix results in a transparent composite. After the pioneering success with the Nylon 6/clay system [1], further advances were reported using other polymer matrices, examples being polyolefins [3], epoxy, [4], polyesters [5], and polyamides [6].

The mechanisms responsible for the property enhancements are generally accepted as associated with the inhibition of polymer molecular motions near the filler surface. Macroscopic measurements of composite properties show that for the nylon/clay system, basic mechanical properties such as modulus, strength and impact strength cease to improve beyond a concentration of 5 wt% (equivalent to 2 vol%) [1]. If we accept the primary role of filler surface, and therefore filler surface area, the existence of a ceiling on property enhancement suggests that an increase in filler volume does not produce a simply related increase in polymer/clay interface area. This is to be expected in systems in which the particles tend to agglomerate, a tendency that would become more difficult to overcome at high concentrations. The importance of filler surface has been demonstrated by Boo et al. [7], where different levels of dispersion were found to correlate with the resultant improvements in mechanical properties. In systems where good dispersion can be maintained, such as that of Zhang et al. [8] consisting of silica spheres in an epoxy matrix, improvements in properties have been observed at up to 14 vol% filler.

In clay systems, platelets of high aspect ratio suitable for nano-reinforcement occur naturally in stacks ('galleries'), so that platelets internal to the stack are isolated from the polymer matrix and relatively little clay surface is exposed to the polymer. Spacing between platelets in this configuration can be measured by X-ray diffraction (e.g. [3,6,9]). More effective reinforcement results from polymer entering the space between the platelets – 'intercalation' – which may be detectable as increased platelet spacing. Complete separation of the platelets – 'exfoliation' – is usually viewed as producing the most effective nanocomposite, though this is an ideal condition and well-dispersed clay nanocomposites will usually contain both exfoliated and intercalated clay [10]. The question of the effectiveness of intercalated galleries a few platelets thick in comparison with fully exfoliated systems is yet to be quantified.

Many workers have observed the distribution and orientation of platelets using transmission electron microscopy (TEM) of thin films. In the Polypropylene (PP)/clay systems that are our particular interest, these observations are a major source of information on the distribution and orientation of the platelets. The degree of exfoliation and the prevalence of intercalated platelets have been investigated, along with their shapes, which show various degrees of curvature [9, 11 – 19]. On the other hand, there have been a number of theoretical studies of nanoreinforcement in which an idealised distribution of platelets is assumed. It is possible to use conventional elastic composite theory to calculate moduli, using the Halpin-Tsai or Mori-Tanaka Equations [20]. In the Halpin-Tsai model, perfectly aligned continuous reinforcement is assumed, and this theory has been exploited for clay nanocomposites by a number of workers [21 – 27]. In the Mori-Tanaka model, an array of ellipsoidal inclusions is assumed, and this theory has been similarly exploited [27,28]. Hbaieeb et al. [28] compared simple models with two- and three-dimensional numerical continuum approaches, and found significant differences.

One of the basic issues of the functioning of nanocomposites is whether, and if so how, different nanoparticles interact. In filled rubbers, it has been suggested that the filler (carbon black or silica) particles can influence one another, so that at sufficient concentration they in effect form a network, resembling an additional molecular network. Klűppel and co-workers have developed a theory of filler networking in a number of publications [29–31], whereby particles up to 1nm apart can interact via the immobilised or glassy layer of material (the 'interphase') that surrounds them. More relevant to work on polymer nanocomposite is that of Shen et al. [32], who studied the rheology of polyamide nanocomposite at various filler levels. They observed a filler concentration threshold, above which their melts exhibited solid-like behaviour, and introduced a 'grafting-percolated'. Monte Carlo simulations with this model suggested that the network would become effective at around 2 vol%, in line with their observations. In a study of PP/clay nanocomposite, Wang et al. [33] discussed the concept of physical 'jamming' of platelets within the three-dimensional polymer network, and concluded from rheological observations that it became effective at around 1 wt%.

The issues that arise here – the distribution and orientation of platelets, their size, shape and aspect ratio, the effectiveness of simplified reinforcement models, and the filler network – can be addressed by numerical simulation. We shall use Finite Element (FE) modelling within a stochastic framework in the development below.

## 2. SEQUENTIAL MULTISCALE MODELING

In the nanocomposite material, phenomena on the molecular scale (angstrom lengthscale), as well on the scale of the platelet configuration (nano/micro lengthscale), affect the overall macroscopic material behaviour. We could not hope to include *all* the underlying degrees of freedom in a macroscopic-sized numerical model, as this would be impossible to realise in practice. Rather, we adopt a multiscale modelling approach – linking a hierarchy of models on the important lengthscales, as indicated in Figure 2 – that reduces the overall number of degrees of freedom by an averaging (homogenisation) process.

There are two basic multiscale modelling approaches: *sequential* (in which a smaller-scale model provides the relevant parameters that form the basis of a larger-scale model, conducted one after the other), and *concurrent* (in which information is interchanged between models on two different scales, executed in parallel). A comprehensive review of multiscale strategies for polymer nanocomposites has recently been presented by Zeng et al. [34]. As the length-scales involved in our multi-scale problem appear to be well separated, we have adopted a sequential ("parameter passing") hierarchical approach, in which structural information is averaged as we step up through the hierarchy of models (Figure 2). On each level, in the sequential approach, the averaged information is *completely* determined before attempting to simulate the level above.

On the intermediate lengthscale, the configuration of platelets is modelled using a statistical representative volume approach. As explained in detail in the following section, an RVE is a "sample" of the heterogeneous material. On this meso-lengthscale, the individual components (platelet and matrix) are treated using continuum mechanics. The FE method is used to determine the "effective" material properties of the nanocomposite (which involves averaging over the sample), which may form the basis of a macroscopic model.

In this work, it is assumed that the platelets are perfectly bonded to the matrix material – a highly oversimplified view. In order to provide a more realistic description, atomistic modelling techniques (such as Monte Carlo or Molecular Dynamics, as discussed in [34]) can be used to provide important information regarding the polymer/platelet interface (e.g. degree of adhesion, slip etc).

Within the framework of the sequential multiscale approach, it is necessary to determine the RVE effective material properties (e.g. the Young's modulus or Stiffness tensor) in advance of the macroscopic analysis. It is the determination of the effective RVE elastic properties that is of central subject of this work.

## 3. REPRESENTATIVE VOLUME ELEMENT

The "*Representative Volume Element*" (RVE) was originally defined by Hill [35] as a sample of model material that is "structurally entirely typical of the whole mixture on average". Since its introduction in 1963, many alternative interpretations of the RVE concept have been suggested: with different ideas about the minimum RVE size and the appropriate boundary conditions, in particular. For a recent discussion regarding RVE definition and size in relation to random heterogeneous material, see [36].

The RVE approach assumes that it is possible to replace the *heterogeneous* (polymer/platelet) material with an equivalent *homogeneous* one, which takes on the average properties of the original heterogeneous material. The volume of model material over which this averaging takes place (the size of the RVE) would ideally be infinitely large. However, in practice (for the purposes of numerical simulation), we define the RVE as a typical sample of model heterogeneous material that contains enough information to reasonably simulate an infinite medium, from which the average ("effective") material properties may be obtained in a meaningful way.

Within the context of the sequential multiscale approach discussed in the previous section, the effective RVE properties are all predetermined, and then applied at the material points of the macroscopic-scale model – e.g. FE analysis. (In the alternative concurrent multiscale approach, an RVE exists or is invoked for each material point involved in the analysis).

Below we define the proper averaging procedure required to determine the effective properties of the nanocomposite; and its use in a statistical interpretation of the above RVE concept. We also describe the details of how to randomly generate RVE geometries that are suitable for FE analysis under periodic boundary conditions

### 3.1. Effective Elastic Material Properties

Consider the case of an RVE subject to only small strains, and linear elasticity (i.e. stress varies linearly with the infinitesimal strain tensor for each individual material). We assume that the continuum approximation is valid at all points within the separate materials, that the polymer and matrix are perfectly bonded, and that there is no voiding or cracking. Note that the use of these assumptions creates a somewhat idealistic material model.

Let $\vartheta$ be a general macroscopic observable quantity. The standard RVE approach defines the "*effective*" value of $\vartheta$ as the volume average of the property over the RVE, that is

$$\langle \vartheta \rangle = \frac{1}{V} \int_V \vartheta(\mathbf{x}) \, dV, \tag{1}$$

where $V$ is the total volume of the RVE, and $\mathbf{x}$ is a position vector. Given a stress field $\sigma_{ij}(x_k)$ and strain field $\varepsilon_{ij}(x_k)$ (where $x_k$ in index notation simply lists components of position), the effective stress and strain components are thus

$$\langle \sigma_{ij} \rangle = \frac{1}{V} \int_V \sigma_{ij}(x_k) \, dV \tag{2}$$

and

$$\langle \varepsilon_{ij} \rangle = \frac{1}{V} \int_V \varepsilon_{ij}(x_k) \, dV. \tag{3}$$

The effective stiffness and compliance tensor elements for the RVE are then defined using these averages via

$$\langle \sigma_{ij} \rangle = C_{ijkl} \langle \varepsilon_{kl} \rangle \tag{4}$$

and

$$\langle \varepsilon_{ij} \rangle = S_{ijkl} \langle \sigma_{kl} \rangle \tag{5}$$

respectively. As discussed above, we can take $C_{ijkl}$ and $S_{ijkl}$ to represent the stiffness and compliance, respectively, of an equivalent homogeneous material – having the same properties as the original heterogeneous material.

The problem of finding the effective elastic material properties, $C_{ijkl}$ or $S_{ijkl}$, requires the stress and strain fields for a typical volume of model composite material under small deformation to be evaluated. In this work we generate RVE geometries with randomly distributed platelets, apply small deformations by displacing the RVE boundaries, and evaluate the resulting stress $\sigma_{ij}(x_k)$ and strain fields $\varepsilon_{ij}(x_k)$ using FE analysis.

### 3.2. Statistical Ensemble

We model the nanocomposite on the intermediate nano- and micro-lengthscale using a *statistical* interpretation of the above RVE approach. This entails the generation of a *statistical ensemble* for the RVE, in a random fashion. A statistical ensemble is a collection of many similar copies (or "realisations") of the system under consideration (a representative sample of heterogeneous nanocomposite material).

Each individual realisation of the ensemble (which we shall refer to as an "*RVE realisation cell*" or "*RVE cell*") must be spatially large enough to reasonably capture the essential physics responsible for the property enhancement (in terms of any long-range order in the stress/strain fields). It follows that the minimum size of the RVE realisation will depend on the details of the platelet configuration: the filling fraction, dispersion, shape, size, material properties etc, as well as on the properties being measured. In practice, the minimum size of the RVE cell was determined, for a given set of parameters, by observing the convergence of measured quantities with increasing RVE cell size. In our results below, RVE cells typically involved of the order 10-100 platelets.

The ensemble must then contain enough RVE cells to give a good statistical representation of the nanocomposite material. The traditional RVE approach involves only a single realisation, which must contain enough information to reasonably represent the infinite medium. In our statistical approach, essentially that same amount of information is contained within the entire ensemble of many smaller RVE cells. Thus, in considering the RVE statistically, we have effectively replaced a *single large* volume with *many small* ones. In general, it is more computationally efficient to simulate many smaller volumes.

The effective stiffness and compliance tensors, defined for the classical RVE in Equations (4) and (5), now involve an average over the ensemble. Let $\bar{\sigma}_{ij}^{(k)}$ and $\bar{\varepsilon}_{ij}^{(k)}$ be the volume-averaged stress and strain components for the $k$th RVE cell in the ensemble. The *ensemble average* of the stress components is given by the arithmetic mean over all RVE cells

$$\left\langle \sigma_{ij} \right\rangle = \frac{1}{N} \sum_{k=1}^{N} \bar{\sigma}_{ij}^{(k)}, \tag{6}$$

and similarly for the strain components

$$\left\langle \varepsilon_{ij} \right\rangle = \frac{1}{N} \sum_{k=1}^{N} \bar{\varepsilon}_{ij}^{(k)}, \tag{7}$$

where $N$ is the number of realisations in the ensemble (sometimes referred to as the "size" of the ensemble – not to be confused with the size of the RVE realisation cell).

These are essentially equivalent to Equations (2) and (3), and so the effective stiffness and compliance tensor elements, Equations (4) and (5), still hold over the ensemble.

A major advantage to the statistical approach is that it naturally provides a good indication of error for measured quantities. If $\vartheta^{(k)}$ is a general quantity measured in the $k$th RVE cell ($k = 1, 2, ..., N$), then in addition to the ensemble average of $\vartheta$

$$\langle \vartheta \rangle = \frac{1}{N} \sum_{k=1}^{N} \vartheta^{(k)} \tag{8}$$

we may also calculate the sample standard deviation in $\vartheta$ over the ensemble

$$s(\vartheta) = \left[ \frac{1}{N-1} \sum_{k=1}^{N} \left( \vartheta^{(k)} - \langle \vartheta \rangle \right)^2 \right]^{\frac{1}{2}}. \tag{9}$$

In practice this allows us to recognise whether the final results have been determined to a reasonable accuracy – increasing the number of realisations if not.

### 3.3. Periodic Boundary Conditions

The most appropriate set of boundary conditions (BC) to impose on an RVE realisation cell is *periodic* BC, as this best simulates the infinite medium limit. With periodic BC, the RVE cell can be thought of as a unit cell, which is repeated in all directions forming an infinite continuous body – thus defining the simulated material, as illustrated in Figure 3. Each copy of the unit cell is referred to as an "image". During simulation, we only need to keep track of information within a single RVE unit cell.

In this work we restrict our attention to the case of a 2D nanocomposite. As can be seen in Figure 3(a), any platelet crossing an edge must reappear at the opposite edge. Indeed, as often happens if a platelet is close to a corner, a single platelet may reappear in several sections.

In practice, periodic BC are easily implemented as follows. Given that the RVE cell (unit cell) is defined in the range $0 \leq x_i \leq L$ in the $i$th direction (in $x$ and $y$), then a value of $x_i$ outside this range (but within an adjacent image $-L \leq x_i \leq 2L$) is adjusted according to

$$x_i \rightarrow \begin{cases} x_i - L, & \text{if } x_i > L \\ x_i + L, & \text{if } x_i < 0 \end{cases}. \tag{10}$$

Also, the periodicity must be taken into account in any operation involving relative distance across the RVE cell: such as checking whether two plate segments intersect, or whether they lie close to one another, which we shall demonstrate later in Section 4. It is often useful for one to imagine, for the $i$th direction, that the two ends of the interval $0 \leq x_i \leq L$ are joined together to form a continuous ring.

A FE mesh must be constructed in a way that allows the FE analysis to be conducted under periodic BC. We accomplish this by insisting that every node appearing on an edge has a corresponding node on the opposite edge. Within the FE calculation, relating all the pairs of opposite boundary-nodes via constraints enforces the periodicity. Consequently, after deformation, the profile of opposite boundaries is identical.

The alternative to the use of periodic BC would be to unrealistically assume reflective symmetry either side of the RVE cell edges, and implement uniformly prescribed BC. This approach would be much simpler to implement – but would introduce undesirable errors near the edges of the RVE cell. The classical RVE approach, which essentially implements a single larger RVE cell, *does* make use of these unrealistic BC. However, the use of reflective BC is justifiable for a *large* realisation: because as the system size increases, the edge-errors incurred as a result of inappropriate BC become less and less significant. The same is not true for our statistical approach, for which the use of many smaller RVE cells makes the use of periodic BC essential.

A similar FE approach to the one used in this work was introduced in 1997 by Gusev [37] to determine the elastic properties using a periodic RVE containing randomly distributed non-overlapping spheres, and later platelets [38]. More recently, the method was applied to polymer/clay nanocomposites by Sheng and Boyce [39] within a multiscale approach.

## 4. GENERATING RVE GEOMETRY

The configuration of platelets observed in the real nanocomposite material very much depends on the exact nature of its preparation. As it would not be feasible to simulate the complex dynamics of the whole preparation stage, we recreate the characteristics of the platelet configuration within the RVE cell in a random fashion.

The characteristics of the platelets (their size, shape, orientation and dispersion) are randomly selected according to given predefined sets of statistical distributions. These distributions may be "extracted" from a large number of images taken of the real nanocomposite – resulting in a statistical recreation of the platelet configuration existing in the real material.

However, in the present study we assume a set of *idealised* statistical distributions (e.g. all platelets having the same length and aspect ratio, having an entirely random orientation, or all being aligned etc.) The distribution of platelets in space is modelled as being completely random, subject to geometrical constraints. The resulting platelet configurations are ideal for investigating the effect each platelet-characteristic has on the overall properties, and thus which are important for property enhancement.

### 4.1. Number of platelets

The "*filling fraction*" is a measure of the amount of platelet material contained in the nanocomposite material. It is most appropriate here to measure the filling fraction by *area* (or volume) rather than by weight. Thus we define the area (volume) filling fraction $f$ as the total area (volume) of platelet divided by the total area (volume) of the combined material.

Given a desired filling fraction, how many platelets must each RVE realisation cell contain? The use of periodic BC implies that the RVE must contain a whole (integer) number of platelets. We are assuming that the platelets are positioned randomly in space (that is, given by a spatial Poisson process), subject to geometrical constraints. Accordingly, the number of platelets in an individual cell should *not* be taken as constant. Rather it should be randomly selected from the Poisson distribution: the probability of finding exactly $n$ ($n = 0,1,2...$) platelets within a particular RVE cell is given by

$$P_n = \frac{\bar{n}^n}{n!}\exp(-\bar{n}).  \quad (11)$$

Here $\bar{n}$ is the average number of platelets expected in a single RVE cell, which is easily calculated from the desired filling fraction, RVE cell size, and the distributions in platelet length and aspect ratio. In order to produce random numbers drawn according to the Poisson distribution (Equation (11)), we used the rejection method given in "Numerical Recipes" [40].

Thus, different RVE cells will contain different numbers of platelets – and some may contain no platelets at all – however, the ensemble average for the number of platelets $\langle n \rangle$ will tend to the desired average $\bar{n}$ as the number of realisations increases.

### 4.2. Generation of platelet configurations

In this work, for the sake of simplicity, we have restricted our attention to the 2D case, corresponding to a thin slice taken through the real 3D nanocomposite material. An example RVE cell is shown in Figure 4. The FE analysis will be performed under conditions of plane stress. Below we give details of the simple rejection-type Monte Carlo algorithm used to construct the RVE cell. Each individual platelet is randomly placed into the RVE cell *sequentially* – the position only being accepted if it "fits in" with the existing platelets (if any), otherwise a new random position is tried.

The imposition of periodic BC is made much easier (and the above mentioned algorithm made more efficient) by initially representing each platelet as a series of straight line segments, as shown in Figure 4(a). The segmentation allows the representation of curved platelets. In our implementation, the degree of segmentation ultimately controls the FE mesh density, as the seeding of the mesh depends on the segmentation. Thus a more complex platelet geometry (which requires a greater segmentation to properly represent it) tends to produce a greater local FE mesh density.

To generate the configuration of platelets within an RVE cell, the following algorithm was used to randomly place each of its $n$ (randomly selected according to Equation (11)) platelets in sequence:
1. Select the current platelet characteristics: length, thickness, curvature and orientation. These will *not* alter.
2. Randomly generate a trial position $(x_{trial}, y_{trial})$ within the RVE cell by generating two unbiased random numbers in the range $[0, L]$.
3. Check for acceptance or rejection – in the attempt to place the current platelet in the trial position $(x_{trial}, y_{trial})$:

a. Reject if any part of the trial platelet overlaps or intersects with an existing platelet.
b. Reject if any part of the trial platelet is within a certain small distance from an existing platelet. This condition takes into account the thickness associated with the platelets.
4. If accepted, add the trial platelet at $(x_{trial}, y_{trial})$, otherwise return to Step 2.

Individual aspects of this algorithm and the complications incurred by the periodic BC are explained in more detail below.

The use of this algorithm *ensures* that the platelet characteristics appear according to the desired statistical distributions. For example, if the selection of platelet orientation $0 \leq \theta \leq 2\pi$ in Step 1 is entirely random, say, then the fact that the platelet is placed by "position-rejection" only, without changing $\theta$ – ensures that the distribution in $\theta$ for all platelets over a sufficiently large number of realisations will indeed be entirely random.

The periodicity of the RVE realisation cell (unit cell) must be taken into account when considering any operation involving distance. This is the case in Step 3 above, where we check to see if two platelets (or rather the segments making up the platelets) intersect, overlap or are within a small distance of one another. An arbitrary point in the unit cell $(x^a, y^a)$, should only "see" the *nearest image* of another point $(x^b, y^b)$. That is, the "appropriate" image (of point $b$) may be in the same cell (as point $a$), or the eight neighbouring cells, as depicted in Figure 3(b), depending on which image is the nearest. This is known as the "minimum image criterion". To test distances in practice, one can either (i) explicitly reproduce the whole cell geometry in the nearest images, (ii) shift any point under consideration (point $b$ above) to each of its nearest images positions, or (iii) make the adjustment

$$\Delta x_i \rightarrow \begin{cases} \Delta x_i - L, & \text{if } \Delta x_i > L/2 \\ \Delta x_i + L, & \text{if } \Delta x_i < L/2 \end{cases}, \quad (12)$$

where $\Delta x_i = x_i^b - x_i^a$ is the $i$ th component ($x$ and $y$) of the distance between two points.

The periodic BC also means that any platelet crossing an edge must reappear at the opposite edge. In our implementation, for simplicity, we insist that the RVE cell is initially a square, with straight rather than jagged edges. Accordingly, if a platelet segment is found to cross an edge, it is "split" at the intersection point, and the remaining length translated (in $x$ or $y$) to the opposite edge (forming an extra segment). As is sometimes the case near a corner of the cell, this splitting process may need to be iterated several times until *all* the original segment length has been placed.

After all the platelets have been placed in the RVE cell, the line segments that we used to represent the platelets are "filled out" to form trapezoids, or rectangles if the platelet is not curved, as shown in Figure 4(c). The reason we disallowed the line segments to be placed within a certain distance from one another, in Step 3(b) of the above algorithm, was to allow enough room for them to be filled out the desired width.

Occasionally the positioning of a platelet causes a very fine mesh to be produced locally. This tends to happen, for example, if the end of the platelet is very close to the edge of the cell, particularly if the platelet is almost parallel to the edge. It is tempting to consider extending the rejection criteria in the above algorithm to disallow such placements. However, if at all possible this should be avoided. As with many Monte Carlo methods, introducing a subtle bias in this way can lead to incorrect results over the ensemble.

As the RVE is filled with more and more platelets, it becomes harder and harder to accommodate further platelets in the spaces left over. This leads to a tendency for platelets to align close to one another almost in parallel. It should be remembered that this is merely an artefact of this simplistic algorithm. Eventually, it becomes impossible to accommodate any more platelets in the RVE cell.

## 5. PERIODIC FINITE ELEMENT MESH

In order to produce a mesh that is suitable for FE analysis with periodic BC, we ensure that every node on the RVE cell boundary has a corresponding node on the opposite edge. These pairs of corresponding nodes, which represent the same point in space, are subject to a "master-salve" type constraint in the FE calculation. Initially (i.e. before any deformation), pairs of corresponding nodes on the left/right faces are geometrically related via

$$x_{\text{right}} = x_{\text{left}} + L \; ; \; y_{\text{right}} = y_{\text{left}}, \qquad (13)$$

and bottom/top pairs via

$$x_{\text{top}} = x_{\text{bottom}} \; ; \; y_{\text{top}} = y_{\text{bottom}} + L, \qquad (14)$$

where $L$ is the initial length of each side of the square RVE cell. The edges of the RVE cell are exactly straight, rather than jagged.

The meshing was performed using the "Triangle" unstructured mesh generator of Shewchuk [41], which produces guaranteed-quality meshes by the Delaunay-based refinement algorithm of [42]. The mesh is "seeded" by supplying node and segment information defining the boundary around each platelet, and the boundary around the whole RVE cell. All the supplied segments reamain in the resulting mesh. The density of seeding sets a target for the local mesh density, as the algorithm attempts to generate a mesh that "fits in" with the seed segments. In order to enforce the quality of the mesh, the algorithm is ordinarily allowed to subdivide the supplied boundary segments, creating extra boundary-nodes (known as "*Steiner points*"). However, we have the ability to suppress the formation of these Steiner points if required.

In addition to the seeding, the size of each triangular element is controlled via a minimum angle constraint; in which nodes are added such that no angle is below a given minimum. We chose this particular mesher for its excellent ability to change from a high mesh-density (required near the platelet ends) to a much lower density (in the spaces between platelets) over a relatively short distance. This tends to reduce the number of nodes in the RVE cell, decreasing the computational effort required to perform the FE analysis.

As the mesher is not intrinsically periodic (i.e. does not observe Equations (12) to (14)), we used the following procedure to produce a mesh that is suitable for FE analysis with period BC:
1. *Seed the boundaries*. Produce a list of nodes and segments, along the boundary of each platelet, and along the RVE cell edges, ensuring that there are matching nodes on opposite faces.
2. *Perform an initial "trial" meshing*. Here we allow the mesher to create Steiner points on the boundaries in order to enforce the mesh quality. In doing so, the nodes of opposite faces of the RVE cell may now no longer match.
3. *Re-seed the boundaries*. In addition to the original seed points from Step 1, we also include pairs of points corresponding with any Steiner points identified in the previous step – such that each and every node on the RVE cell boundary has a corresponding node on the opposite edge.
4. *Construct the final mesh*. A fresh mesh is produced, based on the updated seeding, in which we suppress the creation of Steiner points on the RVE boundary, so that nodes on opposite faces match.
5. *Set the material regions*. The material type (platelet or polymer) is assigned to each element of the mesh (by testing whether an element centroid lies inside a polygon formed from the platelet boundary segments).

The above procedure ensures that a high quality mesh exists over the boundaries. For example, if there is a high density of nodes locally near the edge of the RVE cell, due to the presence of a platelet, then the above procedure will cause the high density of nodes to be carried on over the boundary, to the opposite edge of the cell. Also, the condition that all the nodes appearing on an edge must have corresponding node on the opposite edge is enforced. The geometry of the problem is thus represented by: the position of each node, the element connectivity, the material type of each element, and a list of corresponding RVE cell boundary nodes pairs.

## 6. NUMERICAL SOLUTION PROCESS

As previously described in Section 3.1, the effective elastic properties (stiffness matrix elements) may be deduced by measuring the volume-average stresses that result from the imposition of small applied strains on the RVE cell boundaries. Different components of the stiffness matrix may be "picked out" by individually applying a pure $x$-stretch, a pure y-stretch, or a simple shear strain. Below we describe the solution process, in which the resulting boundary value problem is solved using FE analysis. The measured quantities are then averaged over the whole statistical ensemble of RVE cell realisations. Our implementation of this method is fully automated.

### 6.1. Finite Element analysis of boundary value problem

The set of BC required to apply a pure stretch $\Delta L$ in the $x$-direction, with zero lateral strain ("constant-width $x$-strain") is illustrated in Figure 6. The corner nodes are given prescribed displacements. If $u$ and $v$ are the displacements in the $x$- and $y$-directions, respectively, then

$$u_1 = 0 \quad v_1 = 0$$
$$u_2 = \Delta L \quad v_2 = 0$$
$$u_3 = \Delta L \quad v_3 = 0 \quad , \tag{15}$$
$$u_4 = 0 \quad v_4 = 0$$

where the subscripts label the corner nodes, numbered anti-clockwise starting from lower-left. The other boundary nodes are related by the constraints

$$u_{right} = u_{left} + \Delta L \quad v_{right} = v_{left}$$
$$u_{top} = u_{bottom} \quad v_{top} = v_{bottom} \quad , \tag{16}$$

where $u_{right}$, $v_{right}$, $u_{left}$ and $v_{left}$ refer to the $x$- and $y$-displacements of a right/left corresponding pair of edge-nodes; and $u_{top}$, $v_{top}$, $u_{bottom}$ and $v_{bottom}$ to top/bottom edge-nodes. Thus, the profiles of the top/bottom edges, and also the right/left edges, will always be exactly the same during the deformation.

Within each material region (platelet and matrix), the elastic properties are assumed to be uniform and isotropic. The Young's modulus and Poisson's ratio for the matrix material were taken to be

$$E_{poly} = 0.9 \text{ GPa} \quad v_{poly} = 0.42, \tag{17}$$

being typical values for PP at an ambient temperature of 20ºC. For the Silicate clay platelet material, we used

$$E_{plate} = 160 \text{ GPa} \quad v_{plate} = 0.24, \tag{18}$$

based on the general consensus of reported values over a small subset of the literature. (A broader survey into the elastic moduli of smectite clay platelets has recently been conducted [43], which found the convergence of opinion to be in the range $178 - 265 \text{ GPa}$). The bonding between platelet and matrix is assumed to be perfect.

On imposition of the small applied stain $\varepsilon_{ij}^{applied} \ll 1$, standard FE plane stress analysis was used to solve the boundary value problem, and thus determine the resulting stress field $\sigma_{ij}(x_k)$ within the RVE cell. The field points are in fact the integration points associated with each element. As there is only a single integration point for the 3-noded triangular elements we used, there is only a single set of stress components $\sigma_{ij}(x_k)$ associated with each element. We used the solver capabilities of the commercial FE software package Abaqus/Standard (SIMULIA Corp.) The effective stresses (Equation (2)) associated with this RVE realisation is simply the average over all $M$ elements (both matrix and platelet) contained in the RVE cell

$$\bar{\sigma}_{ij} = \frac{1}{A_{tot}} \sum_{m=1}^{M} A_m \sigma_{ij}^{(m)} \tag{19}$$

where $\sigma_{ij}^{(m)}$ is the stress associated with $m$th element, and

$$A_{tot} = \sum_{m=1}^{M} A_m \tag{20}$$

is the total area of the RVE cell. (If we were to have used elements containing more than one integration point, we would have had to average over the integration area). The components of $\bar{\sigma}_{ij}$ for the RVE cell are recorded (appended to a text file) in order to form the average over all RVE cell realisations that make up the statistical ensemble.

The other types of strains mentioned above (pure *y*-strain and simple shear) may be applied in a similar manner. However, in this work we concentrate on the elastic properties produced for the constant-width stretch in a single direction.

### 6.2. Ensemble averaged elastic properties

As previously discussed in Section 3.2, there are two main reasons for our use of a statistical ensemble of RVE cells, rather than the traditional single-cell approach. The first is that it is often more efficient to solve a large number of small FE problems, rather than one large one (particularly if we were to consider a non-linear problem, where the CPU time rises steeply with the number of nodes in the problem). The second main reason for choosing the statistical approach is that the measured values, such as the elements of the stiffness matrix, have an associated measure of error.

If the average stress measured and recorded for the *k*th member of the ensemble is $\bar{\sigma}_{ij}^{(k)}$, then the effective stress $\langle \sigma_{ij} \rangle$ is the average over the whole ensemble given by Equation (6). For a constant-width *x*-stretch, the effective stress is given by

$$\begin{bmatrix} \langle \sigma_{11} \rangle \\ \langle \sigma_{22} \rangle \\ \langle \sigma_{12} \rangle \end{bmatrix} = \begin{bmatrix} C_{11} & C_{12} & 0 \\ C_{12} & C_{22} & 0 \\ 0 & 0 & C_{66} \end{bmatrix} \begin{bmatrix} \varepsilon_{11}^{\text{applied}} \\ 0 \\ 0 \end{bmatrix}, \qquad (21)$$

where $\varepsilon_{11}^{\text{applied}} = \Delta L / L$, which gives

$$C_{11} = \frac{\langle \sigma_{11} \rangle}{\varepsilon_{11}^{\text{applied}}}, \qquad (22)$$

and also

$$C_{12} = \frac{\langle \sigma_{22} \rangle}{\varepsilon_{11}^{\text{applied}}}. \qquad (23)$$

(The constant-width y-stretch would provide $C_{22}$ and $C_{12}$; and the pure shear $C_{66}$). We are particularly interested in the degree of property enhancement due to the presence of the platelets. Thus, rather than presenting $C_{11}$, it is useful to define the *stiffness enhancement* as $C_{11}/C_{11}^{\text{poly}}$, where $C_{11}^{\text{poly}}$ is the stiffness for the pure polymer matrix material only. For the special case of the platelet orientations being completely random, the nanocomposite as a whole is isotropic, and so the results may be presented in terms of the effective Young's modulus $E = C_{11}(1 - v^2)$ and Poisson's ratio $v = C_{12}/C_{11}$ (plane stress).

### 6.3. Automation

As an ensemble typically consists of over a hundred realisations, the RVE generation, meshing and FE solution process was automated. The main control parameters were the size of the RVE and the applied strain, together with the platelet length, aspect ratio, orientation distribution, curvature, and the number of platelets in the stack (forming an effective particle). There were also some meshing control parameters, such as the seeding density and the minimum angle for the triangulation, which affect the accuracy of the resulting FE analysis. The whole process was broken up into the following steps:
1. Randomly select the number of platelets *n* to be included in this RVE realisation cell, according to Equation (11).
2. Sequentially place each of the *n* platelets in the RVE cell, using the position-rejection Monte Carlo algorithm, as described in Section 4.2, ensuring that platelets crossing the edge of the cell reappear on the opposite side.
3. Mesh the geometry using the two-pass scheme described in Section 5, which is based on the Triangles mesher. For every edge node there is a corresponding node on the opposite side, geometrically related via Equations (13) or (14).
4. Make an input (.inp) file for the FE solver Abaqus. This is a text file that contains all the information required for the analysis: the node coordinates, element connectivity, material type of each element, a list of edge-node pairs, and the boundary strain.
5. Solve the FE problem by executing Abaqus/Standard.
6. Perform averages over the RVE cell elements using Equations (19) to (20): extracting the required data from the Abaqus output files (.dat and .fil). Append a line containing the results for this realisation to an "ensemble results" text file, for later analysis.

After repeating the procedure for all *N* members of the ensemble, the ensemble averages were then calculated, Equations (6) to (9), using the values recorded in the final step above. If the desired accuracy in the ensemble average (as indicated by its standard deviation) was not achieved, then more realisations were performed and added to the ensemble, until the required accuracy was reached. The effective elastic properties were deduced from the average stresses as explained in Section (6.2).

Each process in the scheme was implemented separately: with text file input and text file output. This is extremely useful for testing and debugging purposes. There are various ways in which one could automate the process. We wrote a menu driven "controlling" program that called each program – by passing a string to the command-line of the operating system – either singly or iteratively. Thus, an entire ensemble of RVE realisations, or indeed a set of ensembles, could be "setup" and executed automatically.

## 7. ELASTIC RVE NUMERICAL RESULTS

Continuing with the example geometry we have been using thus far, Figure 7 shows the final (deformed) geometry and the stress field after undergoing a 1% constant-width *x*-strain. Here, the periodic boundary value problem was solved using FE analysis under the condition of plane stress, as explained in Section (6.1),. Note that this cell is much smaller than the proper RVE realisation cells used later: the size was chosen simply in order to show the various features more clearly. One can see that, due to the imposed periodic boundary constraints given by Equation (16), the profiles of opposite edges are indeed identical. Also, the extent to which the deformation has affected the edge profiles, in this example, demonstrates that significant edge-errors would have been incurred had we imposed uniform displacement BC (where the edges remain straight), rather than the more appropriate periodic BC.

As can be seen in Figure 7(a), upon deformation, the stress is mainly taken up by the platelets: particularly those that are close to being aligned parallel with the stretch-direction. In fact, the magnitude of the stress in the platelets is up to an order of magnitude greater than that in the polymer. Note that platelets orientated almost perpendicular to the stretch direction carry *much* less stress compared with those parallel to it.

For platelets that are in relative isolation, the largest stress appears to occur towards the centre of the platelet. As shown quantitatively by Sheng and Boyce et al.[39], this indicates that the load-transfer from matrix to platelet is mainly through interface shear stress, giving rise to the classical "shear-lag" effect [44] along the length of the platelet. One can also clearly see that the stress in the platelets is affected by the presence of other nearby platelets (platelet-platelet interaction).

In Figure 7(b), the maximum contour value has been reduced in order to reveal the stress field that exists in the polymer matrix. One can see that the stress in the polymer tends to be lowest near the long edge of the platelets (particular towards the middle), and relatively high close to the platelet "tips". This important feature results from the fact that a platelet is much less strained than polymer material away from its immediate vicinity. As there is perfect bonding between platelet and polymer, the presence of the platelet constrains polymer that is close to its long edge, causing large deformation to be produced in the polymer close to its tips.

The high areas of stress that appear to "emanate" from a platelet tip tend to be greater when another platelet is nearby, in fact causing the nearby platelet to bend. Note that this interaction is particularly large when the platelets are pointing directly toward one another, and when they are parallel with the stretch direction. Thus, there appears to be long-range order in the stress field that extends beyond the length of a single platelet – possibly forming some kind of network.

The cell shown in Figure 7 is clearly not large enough to properly represent the problem: as there may be a significant interaction between the "head" end of a platelet and its *own* "tail" end, across the periodic boundary. We must choose a cell size $L$ that is sufficiently large, no only to render this periodic artefact negligible, but also to reasonably encompass the relevant long-range order in the stress field. The long-range order will depend on the various platelet characteristics: particularly the platelet filling fraction.

The "standard" RVE realisation cell used in the work below contained platelets of aspect ratio 100:1 to a filling fraction of 4% (by area). The standard elastic constants for the matrix and platelet materials are given by equations (17) and (18), and the applied strain was always 1%. Both straight and curved platelets were generally constructed using 10 segments (the corners of which acted as seed points), and meshed such that no angle was below 30º.

The method used to determine a reasonable value to take for $L$ was to observe the convergence of the measured stress with increasing $L$, keeping the platelets-characteristics fixed. Two such curves are shown in Figure 8, using the "standard" parameters above, for randomly aligned platelets, as in Figure 9, and for platelets aligned with the $x$-direction, as in Figure 15(a). Each point represents the average of the stress $\bar{\sigma}_{11}$ taken over all the elements in the RVE cell, averaged over an ensemble of between 25 and 250 such realisation cells. (The larger the value of *L*, the fewer realisations were required to produce an ensemble average to a given accuracy, as each realisation contains more information). As can be seen from Figure 8, the curves converge quickly to their respective infinite cell limit. We took the RVE cell size *L* to be 5 times the standard platelet length. Depending on the filling fraction, each RVE cell realisation contained of the order 10-100 platelets. The CPU time required to solve the problem for each realisation was typically less than 1 minute on a modern PC. In this work, averages were taken over ensembles of between 25 and 250 realisations

In reality, polymer/clay nanocomposites typically consists of a mixture of fully exfoliated single silicate layer platelets, and partially exfoliated particles consisting of a number of silicate layers together (multi-layer stacks of platelets). We begin below by considering the case where all platelets are fully exfoliated, and examine the way in which the various platelet characteristics affect the effective elastics properties. In particular, we investigate the effect of altering the curvature of the platelets, and also the platelet orientation. Finally, we construct RVE geometries containing multi-layer stacks of platelets, enabling us to study how the degree of exfoliation affects the effective properties of the nanocomposite material.

**7.1. Fully exfoliated straight platelets**

Consider an RVE containing only randomly-orientated single-layered straight platelets only. This is the "ideal" case of full exfoliation and complete dispersion, with straight platelets. An example RVE realisation cell for such a configuration is shown in Figure 9, where the platelet aspect ratio (length/width) is 100:1, the length of the RVE cell is 5 times the platelet length, and the filling fraction is 4% by area. Below we will examine the way in which the various platelet characteristics affect the overall effective stiffness in the *x*-direction $C_{11}$, by applying of a 1% constant-width *x*-stretch. The results for $C_{11}$ will be presented in units of its value for pure polymer $C_{11}(\text{pure}) = 1.09273$. Thus, a "stiffness enhancement" $C_{11}/C_{11}(\text{pure}) > 1$ indicates an improvement of mechanical properties in the *x*-direction due to the presence of the platelets; and $C_{11}/C_{11}(\text{pure}) < 1$ would indicate a reduction.

Firstly, let us see how the relative platelet stiffness $E_{\text{plate}}/E_{\text{poly}}$ affects the overall effective stiffness of the nanocomposite, where $E_{\text{poly}}$ is given Equation (17). Figure 10 shows the stiffness enhancement $C_{11}/C_{11}(\text{pure})$ for values of platelet stiffness up to 600 times that of the polymer, whist keeping the aspect ratio and filling fraction constant at their standard values of 100:1 and 4% respectively. Each point represents an average over an ensemble of 30-40 realisations. As one can see, the stiffness initially rises sharply, and then levels out towards the infinite platelet-stiffness limit. If the value of $E_{\text{plate}}/E_{\text{poly}} \approx 200$ for the real polymer/clay nanocomposite, then we can see the stiffness enhancement is already fairly close to the maximum possible value. That is, using a stiffer material in the platelet would not lead to a significant improvement in the mechanical properties.

Now consider the way in which the aspect ratio (defined as the platelet length divided by its width) of the platelets changes the overall RVE stiffness. The variation of stiffness enhancement with aspect ratio is shown in Figure 11, keeping the filling fraction constant. Thus, as the value taken for the aspect ratio increases, the RVE contains a *larger* number of *longer/thinner* platelets. As can be seen, for aspect ratios above 100, the increase in stiffness enhancement appears to be linear up to the aspect ratio shown in the plot.

In Figure 12, we present the variation of stiffness enhancement with platelet area filling fraction $f$, holding the platelet aspect ratio constant at 100:1. Each point represents an average over 25-100 realisations, where cell configurations involving fewer platelets require a larger number of realisations (there is grater variation in the number of platelets selected from the Poisson distribution). Here we compare the case of the randomly orientated and randomly positioned platelets; with the case of platelets that are all aligned in the *x*-direction, and are randomly positioned. In both cases there is an increase in stiffness enhancement that is approximately linear with increasing $f$. (In the real nanocomposite, we would expect the increase to be less rapid with increasing $f$ due to the effects of increased particle agglomeration). We observe from Figure 12 that, for all filling fractions, the stiffness enhancement for the aligned platelets is more than *double* that for the randomly aligned platelets. This dependence on platelet orientation is investigated in greater detail below.

**7.2. Effect of platelet orientation**

The orientation of the platelets in the real nanocomposite appears to be approximately random. However, during processing (e.g. extrusion), the platelets tend to become preferentially orientated in the processing direction, as can be seen from the images in Figure 1. We have already observed above, from Figure 12, that the stiffness enhancement for the case of perfectly aligned platelets is more than double that for the case of completely random platelets, over all filling fractions.

Staying with fully exfoliated (single layer) straight platelets that are randomly positioned in space, we now investigate the way in which the *degree* of platelet orientation effects the overall elastic properties. Accordingly, two separate ways of gradually altering the platelet orientation distribution are considered below. All data points presented were averaged over ensembles of 25 RVE cell realisations, each containing platelets of aspect ratio 100:1 and filled to 4% by area.

Firstly, consider an RVE platelet configuration in which *every* platelet is given the *same* angle of tilt $\pm\theta$ to the *x*-direction. Figure 13 shows the stiffness enhancement for tilt angles ranging from $\theta = 0$ (all platelets perfectly aligned with the direction of applied strain) to $\theta = 90°$ (all platelets are aligned in the direction normal to the applied strain). Thus, this plot encompasses the two furthest possible extremes in platelet alignment. As can be seen from the figure, the stiffness enhancement has a Gaussian-like shape: with its maximum possible value for perfect alignment at $\theta = 0$, reducing to its minimum possible value for the case of perpendicular alignment (for which there is virtually no enhancement).

Now consider the somewhat more physically realistic RVE configuration in which the platelet orientation $\theta$ is uniformly random up to a maximum tilt angle of $\pm\theta_{max}$ from the x-direction. In Figure 14 we present the stiffness enhancement from $\theta_{max} = 0$ (all platelets fully aligned with the direction of applied strain) to $\theta_{max} = 90°$ (completely random platelet orientation). Here we transverse between the two main orientation distributions of interest. Configurations corresponding to different values of $\theta_{max}$ are similar to those one might expect for various "degrees" of processing: from the randomly orientated state on exfoliation with no processing ($\theta_{max} = 90°$), through to the perfectly aligned state representing "extreme" processing ($\theta_{max} = 0$). As can be seen, there is an approximately linear reduction in the stiffness enhancement with increasing $\theta_{max}$.

Clearly we may conclude that the effective properties in the direction of any preferential platelet orientation are enhanced considerably. The load transfer from polymer to platelet is more effective for platelets that are close to being aligned with the direction of applied strain, leading to a more stress being taken up by the platelets, and therefore greater reinforcement. There also appears to be a complicated platelet-platelet interaction taking place. In the transverse direction to any platelet alignment, the mechanical properties are likely to be reduced.

### 7.3. Curved Platelets

On exfoliation, the platelets in the real unstretched nanocomposite material tend to be curved, rather than straight, as can be clearly seen in Figure 1(a). We now investigate how platelet curvature might affect the overall elastic properties: by considering RVE geometries containing randomly positioned platelets that *all* have the *same* constant curvature.

The same construction method was used to generate the curved platelets as for the straight platelets, discussed in Section 4.2. As a platelet is made up of a series of straight-line segments – causing the platelet to have constant curvature is simple a matter of applying a small offset angle between each segment.

We compared RVE configurations containing either straight or constant-curvature platelets. In both cases, the platelets consisted on 10 segments, were of aspect ratio 100:1, and were filled to an area fraction of 4%. Figure 15 shows a comparison of RVE cells for straight platelets aligned with the x-direction, and constant-curvature platelets aligned in the x-direction (the direction being that of the line joining its end-points). Notice that we appear to get stress "shielding" in the polymer where it is "cupped" by the curvature of the platelet (the polymer is constrained). We also considered straight and constant-curvature configurations both with completely random platelet alignment.

The stiffness enhancement for each of the 4 cases described above is plotted against filling fraction in Figure 16, averaged over 25-60 realisations. It shows that there is little difference in stiffness enhancement between straight and curved platelets if they are randomly orientated. However, there is a significant difference between straight and curved platelets if they are aligned with the direction in which the stain is applied.

The difference in overall stiffness enhancement for the aligned case is mainly due to the fact that the curved platelets can not achieve the maximum polymer-to-platelet load-transfer which exists for the straight platelets. This is because only a small proportion of a curved platelet is in good alignment with the direction of applied strain (also one interface is in fact shielded), whereas in *whole* of a straight platelet is aligned. (We have already observed that the load transfer is highly dependant on orientation). The curvature may also cause a difference in the platelet-platelet interaction.

### 7.4. Multi-layer stacks of intercalated platelets

In general, real polymer/clay nanocomposites are *not* in a state of complete exfoliation (in which *all* the clay is fully separated into single layered platelets and dispersed in the polymer matrix). Rather, these materials typically consist of a mixture of *some* fully exfoliated platelets together with stacks of intercalated platelets a few layers thick. An intercalated stack is one in which the inter-layer gallery material has been replaced by the polymer matrix material. Below we investigate the way in which the number of layers in these stacks alters the property enhancement.

The same position-rejection method previously used for single-layer platelets, discussed in Section 4.2, is easily extended to generate multi-layer "effective" particles. Given an orientation $\theta$ and the randomly generated trial position $(x_{\text{trial}}, y_{\text{trial}})$ as before, then the start position of the $k$th layer in the stack is $(x_k, y_k)$ where

$$x_k = x_{\text{trial}} + d(k-1)\cos(\theta + \pi/2)$$
$$y_k = y_{\text{trial}} + d(k-1)\sin(\theta + \pi/2)$$
(24)

which produces a stack of parallel platelets, separated by the inter-layer spacing $d$, with ends aligned perpendicular to the length. During the construction of the stack, if any part is found to overlap or intersect with existing platelets (already successfully placed in the RVE cell) then this trial is rejected, and another $(x_{\text{trial}}, y_{\text{trial}})$ tried. Thus we may construct configurations of multi-layer stacks which tend to the desired statistical distributions (in $\theta$ for example) as the number of realisations increases. In all cases, the aspect ratio of platelets making up the stacks was 100:1, and the inter-layer spacing $d$ was taken to be 4 times the platelet width (thus the thickness of the polymer separating the layers was taken to 3 times the platelet width).

We constructed RVE configurations containing multi-layer stacks *all* of which had the *same* number of layers $n_{\text{layer}}$. The polymer material in the inter-layer regions was assumed to have the same elastic properties as the polymer surrounding the stacks. Also, perfect bonding is still assumed at the polymer/platelet interface. An example RVE geometry containing 5-layer stacks of platelets filled to an area fraction of 4% is shown in Figure 18. We observe that the stacks appear to act as "effective particles".

We considered the stiffness enhancement both for randomly aligned stacks (shown in Figure 18), and for stacks aligned with the applied-strain direction (shown in figure 19), for $n_{layer}$=1,2,3,4, and 5 at various area filling fractions $f$ (of 1%, 2%, 3% and 4%). Each point in the plots represents an ensemble average over 250 realisations (thus the data presented in Figures 18 and 19 was produced using 10,000 separate RVE cell realisations). On can see that the plots for random- and aligned-orientation are very similar in shape, but the magnitude of the stiffness enhancement is larger for the case of the aligned stacks. This indicates that the multi-layer stacks have a similar orientation-dependence as we found for the single-layer platelets.

It is immediately obvious that, for all the $n_{layer}$ values studied, the stiffness enhancement increases with increasing filling fraction. However, one can also see that, for a given value the filling fraction $f$, the stiffness enhancement decreases and also *levels off* as the value of $n_{layer}$ increases. This levelling off effect is mainly due to the fact that the platelets internal to the stack tend to be shielded from the stress, making them less effective than fully exfoliated single platelets.

Clearly, it would be desirable to produce a nanocomposite containing mainly single fully-exfoliated platelets at high filling fractions. In practice, the increase in agglomeration at higher filling fractions (which we did not consider in our model) prevents this. However, from the results above, we observe that stacks of 2 and 3 layers still provide reasonable reinforcement at higher filling fractions.

## 8. CONCLUSIONS

The improved mechanical properties in polymer/clay nanocomposites are associated with particle geometries of high aspect ratio and the resulting high interfacial area per unit volume. In this work, an elastic Finite Element analysis of idealised clay platelet configurations was carried out: identifying which platelet characteristics are important in producing the property enhancement.

We used a *statistical* interpretation of the Representative Volume Element approach, which averages the material properties over a large number of relatively small periodic realisation cells of the model system. The advantage in using many small realisation cells rather than a single large cell (the traditional approach) is that the solution process is computationally more efficient, and it provides a good indication of error in the final results. A "position-rejection" algorithm was used to randomly place the clay platelets within each realisation cell *sequentially*, such that they do not overlap or intersect. This simple algorithm guarantees that the desired statistical distributions – in platelet filling fraction, size, shape, position and orientation – are produced over the ensemble. The effective elastic material properties were deduced by measuring the response to a small applied strain for each cell – using standard Finite Element analysis to solve the periodic boundary value problem – and then averaging over the entire ensemble. In our implementation, the whole process was automated.

On application of the applied strain, the resulting stress is mainly taken up by the platelets, which thus act to reinforce the material. The axial stress in the platelets is highest towards the platelet centre, indicating that the load-transfer from polymer to platelet is primarily through interface shear stress (shear-lag effect). We observed that the presence of a platelet tends to constrain the polymer along its long edge ("stress shielding"), and produces high deformation near the platelet ends (giving the appearance of high stress "emanating" from platelet tips). The stress fields in and around the platelets were clearly affected by the presence of other nearby platelets, which may possibly give rise to some kind of network in the real material.

For the "ideal" case of full exfoliation, in which the platelets are all separate and randomly distributed, we found that the stiffness enhancement (ratio of the stiffness of the nanocomposite to that of the pure polymer) increased with increasing platelet aspect ratio, platelet stiffness, and filling fraction. The effect of platelet orientation is of particular importance (platelets tend to become preferentially aligned on processing). By allowing platelets to be randomly orientated only up to a maximum tilt angle, we found that the stiffness enhancement reduced approximately linearly as the degree of alignment was reduced: from all perfectly aligned with the stretch direction, to completely random. The load transfer from polymer to platelet is more effective for platelets are close to being aligned with the stretch direction. We also studies configurations of platelets with constant-curvature (platelets tend to be curved on exfoliation). We found little difference in stiffness enhancement between curved and straight platelets with random orientation; however there was a significant difference between curved and straight platelets aligned with the stretch-direction. This is because curved platelets only have small proportion of their length orientated close to the stretch-direction.

We also considered the case of multilayer stacks of intercalated platelets: comparing configurations containing only stacks of a particular number of layers ($n_{layer}$=1,2,3,4, and 5), at various filling fractions, for both aligned and random stack orientations. For all the cases studies, the stiffness enhancement increased with increasing filling fraction. Also, the stiffness enhancement tends to decrease and *level off* as the number of layers in the stack increases. This is because the platelets internal to the stack are shielded from the stress by the outermost platelets, making them less effective. We observe that the stiffness enhancement has a similar value for stacks containing 4 and 5 layers for a given filling fraction and alignment. The results show that the greatest enhancement occurs in the case of full exfoliation. However, we still see reasonable stiffness enhancement for stacks of 2 and 3 layers at higher filling fractions.

**ACKNOWLEDGMENTS**

We acknowledge EPSRC (UK) for funding this project.

**REFERENCES**

1. Okada A and Usuki A (2006) Twenty Years of Polymer-Clay Nanocomposites *Macromol. Mater. Eng.* **291**: 1449–1476.
2. Hussain F, Jojjati M, Okamoto M and Gorga E R (2006) Review article: Polymer-matrix Nanocomposites, Processing, Manufacturing, and Application: An Overview *J. Compos. Mater.* **40**: 1511–1575.

3. Gopakumar T G, Lee J A, Kontopoulou M and Parent J S (2002) Influence of clay exfoliation on the physical properties of montmorillonite/polyethylene composites *Polymer* **43**: 5483–5491.
4. Kornmann X, Thomann R, Műlhaupt R, Finter J and Berglund LA (2002) High Performance Epoxy-Layered Silicate Nanocomposites *Polym. Eng. Sci.* **42**: 1815–1826.
5 Ke YC, Yang ZB and Zhu CF (2002) Investigation of properties, nanostructure, and distribution in controlled polyester polymerization with layered silicate *J. Appl. Polym. Sci*. **85**: 2677–269.
6. McNally A, Raymond Murphy W, Lew C Y, Turner R J and Brennan G P (2003) Polyamide-12 layered silicate nanocomposites by melt blending *Polymer* **44**: 2761–2772.
7 Boo W J, Sn L, Liu J, Moghbelli E, Clearfield A, Sue H-J, Pham P and Vergese N (2007) *J. Polym. Sci., Part B: Polym. Phy.* **45**: 1459–1469.
8. Zhang H, Zhang Z, Friedrich K and Eger C (2006) Property improvements of in situ epoxy nanocomposites with reduced interparticle distance at high nanosilica content *Acta Materiala* **54**: 1833–1842.
9. Chiu F-C, Lai S-M, Chen J-W and Chu P-H (2004) Combined Effects of Clay Modifications and Compatibilizers on the Formation and Physical Properties of Melt-Mixed polypropylene/clay nanocomposites *J. Polym. Sci., Part B: Polym. Phys.* **42**: 4139–4150.
10. Morgan A B, Gilman J W (2003) Characterization of polymer-layered silicate (clay) nanocomposites by transmission electron microscopy and X-ray diffraction: A comparative study *J. Appl. Polym. Sci*. **87**: 1329–1338.
11 Liu X and Wu Q (2001) PP/clay nanocomposites prepared by grafting-melt intercalation *Polymer* **42**: 10013–10019.
12 Pereira de Abreu D A, Paseiro Losada P, Angulo I and Cruz J M (2007) Development of new polyolefin films with nanoclays for application in food packaging *Eur. Polym. J.* **43**: 2229–2243
13 Lee E C, Mielewski D F and Baird R J (2004) Exfoliation and dispersion enhancement in polypropylene nanocomposites by in-situ melt phase ultrasonication *Polym. Eng. Sci*, **44**: 1773–1782.
14 Naderi G, Pierre G. Lafleur P G and Dubois C (2007) Microstructure-properties correlations in dynamically vulcanized nanocomposite thermoplastic elastomers based on PP/EPDM *Polym. Eng. Sci*. **47**: 207–217.
15 Shi D, Wei Y, Li R K Y, Ke Z and Yin J (2007) An investigation on the dispersion of montmorillonite (MMT) primary particles in PP matrix *Eur. Polym. J.* **43**: 3250–3257.
16 Yuan Q and Misra R D K (2006) Impact fracture behavior of clay–reinforced polypropylene nanocomposites *Polymer* **47**: 4421–4433.
17 Maiti P, Nam P H, Okamoto M and Kotaka T (2002) The effect of crystallization on the structure and morphology of polypropylene/clay nanocomposites *Polym. Eng. Sci*. **42:** 1864–1871.
18 Yuan Q, Awate S and Misra R D K (2006) Nonisothermal crystallization behavior of polypropylene–clay nanocomposites *Eur. Polym. J.* **42**: 1994–2003.
19 Kanny K and Moodley V K (2007) Characterisation of polypropylene nanocomposite structures Journal of Engineering Materials and Technology **129**: 105–112.
20 Ward I M and Sweeney J (2004) in *An introduction to the mechanical properties of solid polymers (2$^{nd}$. Edn.)*, Wiley, Chichester.

21 Stretz H A, Paul DR, Li R, Keskkula H and Cassidy P E (2005) Intercalation and exfoliation relationships in melt-processed poly(styrene-co-acrylonitrile)/montmorillonite nanocomposites *Polymer* **46**: 2621–2637.

22 Vlasveld D P N, Groenewold J, Bersee H E N, Mendes E and Picken S J (2005) Analysis of the modulus of polyamide-6 silicate nanocomposites using moisture controlled variation of the matrix properties *Polymer* **46**: 6102–6113.

23 Osman M A, Rupp J E P and Suter U W (2005) Tensile properties of polyethylene-layered silicate nanocomposites *Polymer* **46**: 1653–1660.

24 Zhong Y, Janes D, Zheng Y, Hetzer M and De Kee D (2007) Mechanical and oxygen barrier properties of organoclay-polyethylene nanocomposite films *Polym. Eng. Sci*. **47**: 1101–1107.

25 Rao Y Q (2007) Gelatin-clay nanocomposites of improved properties *Polymer* **48**: 5369–5375.

26 Kalaitzidou K, Fukushima H, Miyagawa H and Drzal L T (2007) Flexural and tensile moduli of polypropylene nanocomposites and comparison of expermental data to Halpin-Tsai and Tandon-Wang models *Polym. Eng. Sci*. **47:** 1796–1803.

27 Fornes T D and Paul D R (2003) Modeling properties of nylon 6/clay nanocomposites using composite theories *Polymer* **44:** 4993–5013.

28 Hbaieb K, Wang Q X, Chia Y H J and Cotterell B (2007) Modelling stiffness of polymer/clay nanocomposites *Polymer* **48**: 901–909.

29 Heinrich G, Klűppel M and Vilgis T (2007) in *Physical Properties of Polymer Handbook (2$^{nd}$ Edn.)* Mark EJ (Ed.) Chapter 36, Springer.

30 Heinrich G and Klűppel M (2002) Recent advances in the theory of filler networking in elastomers *Adv. Polym. Sci.* **160**: 1–44.

31 Meier J G, Mani J W and Klűppel M (2007) Analysis of carbon black networking in elastomers by dielectric spectroscopy *Phys. Rev*. *B* **75**: 054202-1-054202-10.

32 Shen L, Lin Y, Du Q, Zhong W and Yang Y (2005) Preparation and rheology of polyamide-6/attapulgite nanocomposites and studies on their percolated structure *Polymer* **46**: 5758–5766.

33 Wang K, Liang S, Deng J, Yang H, Zhang Q, Fu Q, Dong X, Wang D and Han C C (2006) The role of the clay network on molecular chain mobility and relaxation in polypropylene/organoclay nanocomposites *Polymer* **47**: 7131–7144.

34 Zeng Q H, Yu A B and Lu G Q (2008) Multiscale modeling and simulation of polymer nanocomposites *Prog. Polym. Sci.* **33**: 191–269.

35 Hill R (1963) Elastic properties of reinforced solids: some theoretical principles *J Mech Phys Solids* **11**: 357–372.

36 Gitman I M, Askes H and Sluys L J (2007) Representative volume: Existence and size determination *Engineering Fracture Mechanics* **74**: 2518–2534.

37 Gusev A A (1997) Representative volume element size for elastic composites: a numerical study *Journal of the Mechanics and Physics of Solids* **45**: 1449–1459.

38 Gusev A A (2001) Numerical Identification of the Potential of Whisker and Platelet Filled Polymers *Macromolecules* **34**(9): 3081–3093.

39 Sheng N, Boyce M C, Parks D M, Rutledge G C, Abes J J and Cohen R E (2004) Multiscale Micromechanical Modeling of Polymer/Clay Nanocomposites and the Effective Clay Particle *Polymer* **45**(2): 487–506.

40 Press W H, Flannery B P, Teukolsky S A and Vetterling W T (1986) in *Numerical recipes, the art of scientific computing* Chapter 7, Cambridge University Press, London.

41 Shewchuk J R (2005) Triangle, A Two-Dimensional Quality Mesh Generator and Delaunay Triangulator, version 1.6

*http://www.cs.cmu.edu/~quake/triangle.html.*
42 Ruppert J (1995) A Delaunay Refinement Algorithm for Quality 2-Dimensional Mesh Generation. *Journal of Algorithms* **18**(3): 548–585.
43 Chen B and Evans J R G (2006) Elastic moduli of clay platelets *Scripta Materialia* **54**: 1581–1585.
44 Cox H L (1952) The Elasticity and Strength of Paper and Other Fibrous Materials *Brit J Appl Phys* **3**: 72–79.

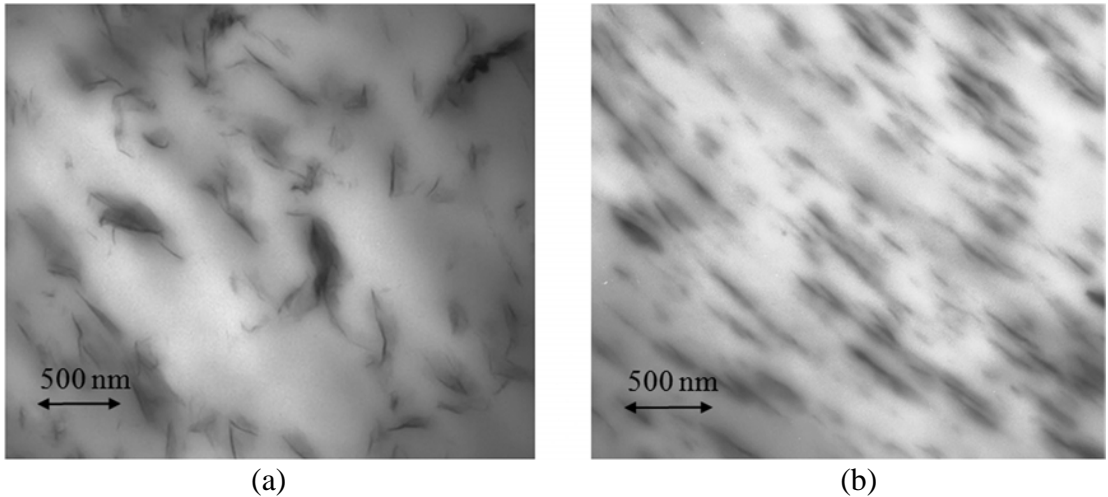

(a)                                                 (b)

Figure 1: Transmission electron microscopy images of Polypropylene filled with nano-clay to 5% by weight (courtesy of Queen's University Belfast). (a) The combination of exfoliated and intercalated clay platelets is evident. (b) After stretching the platelets appear to be less curved and are aligned in the stretch-direction.

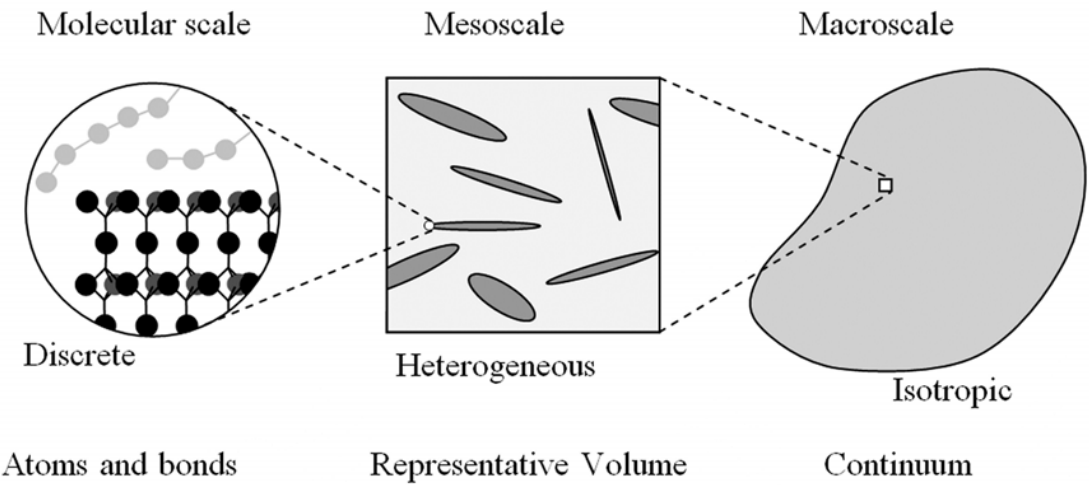

Figure 2: The relevant lengthscales involved in the hierarchical multiscale modelling of nanocomposite materials.

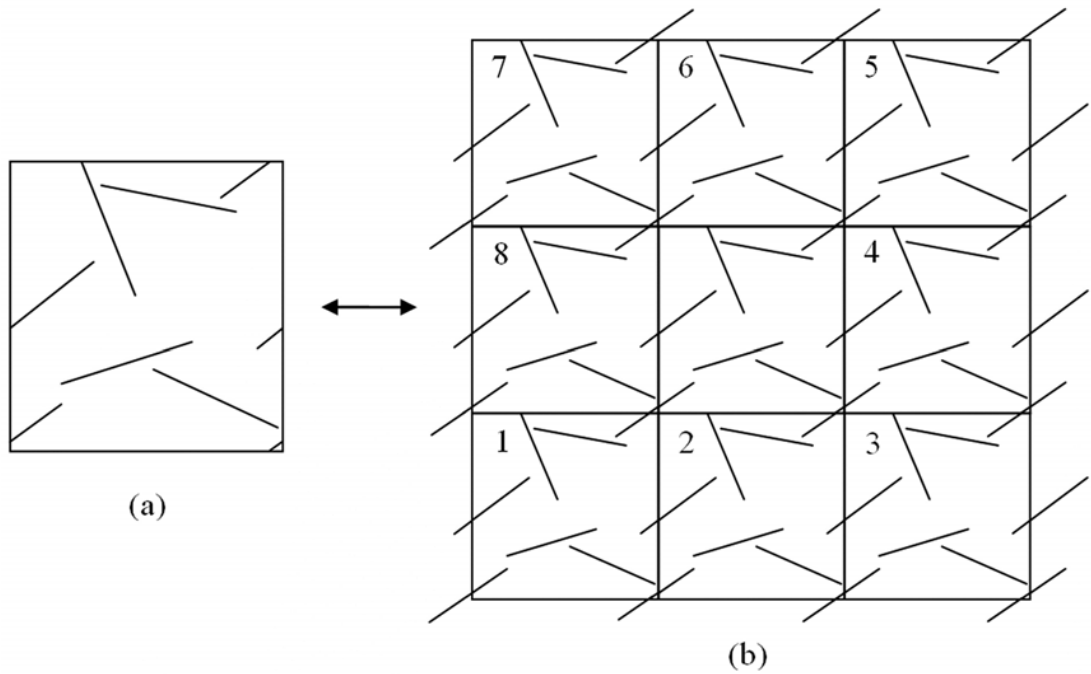

Figure 3: Unit cell with periodic boundary conditions. (a) Any platelet passing through an edge of the unit cell reappears on the opposite edge. Thus the unit cell may be repeated in all directions to form an infinite continuous body. In (b), we show the (central) unit cell with its nearest "images" (labelled 1-8). A platelet (in the central cell) will *only* "see" the *nearest* image of any other platelet (within the 9 cells shown).

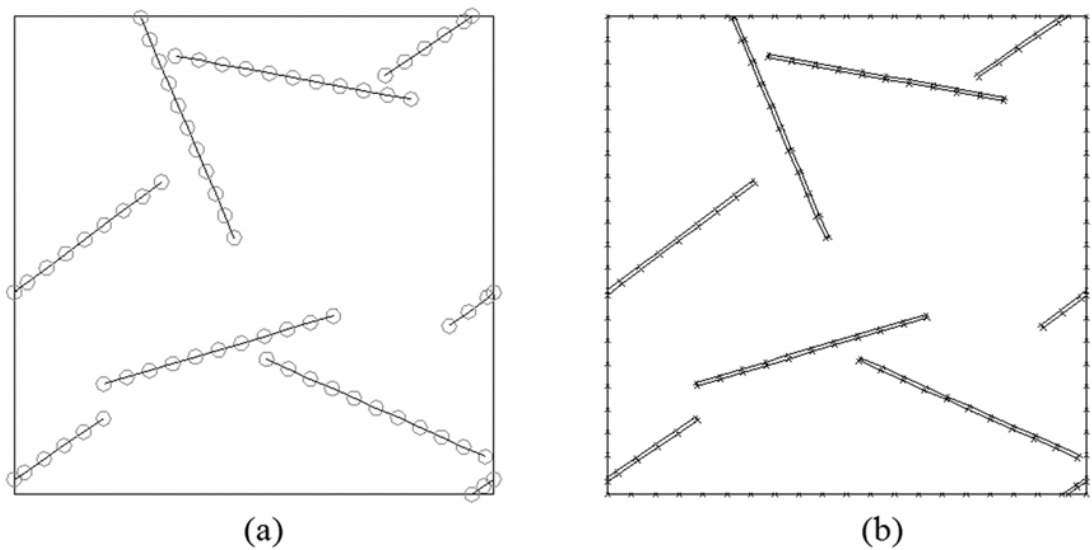

Figure 4: Construction of a single RVE realisation cell geometry, representing a thin slice through the real 3D nanocomposite. (a) The platelets are initially represented by 1D straight line segments, and then (b) "filled out" to the desired width.

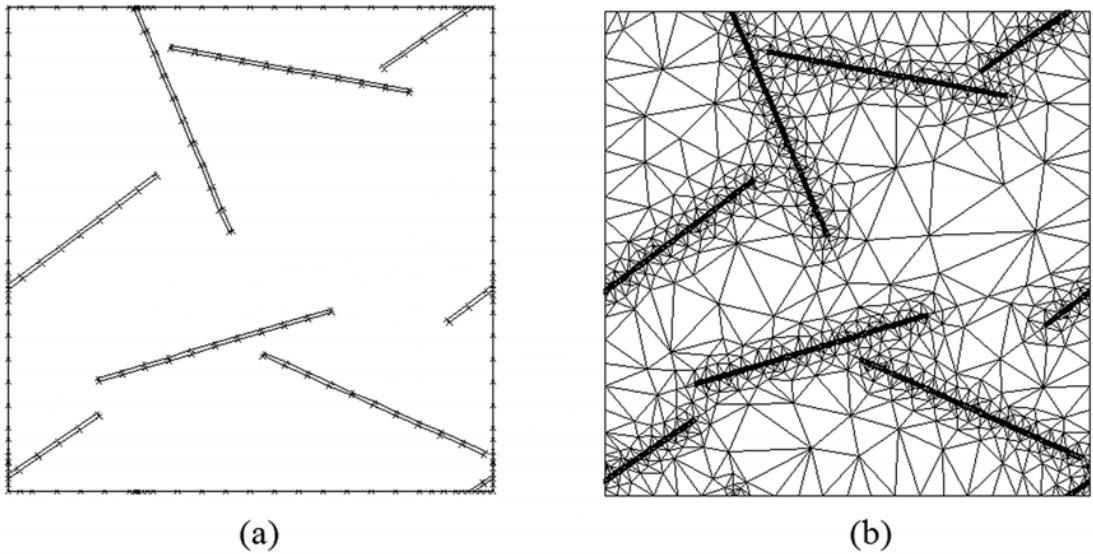

Figure 5: (a) The mesh is seeded such that every node on the left edge has a corresponding node on the right edge. The same is true for bottom/top nodes. An initial pass through the mesher provides appropriate seeding in regions of high mesh density. (b) The resulting mesh has highest density in and near to the platelets. The seeding ensures that regions of high mesh density pass through the boundaries to the opposite face.

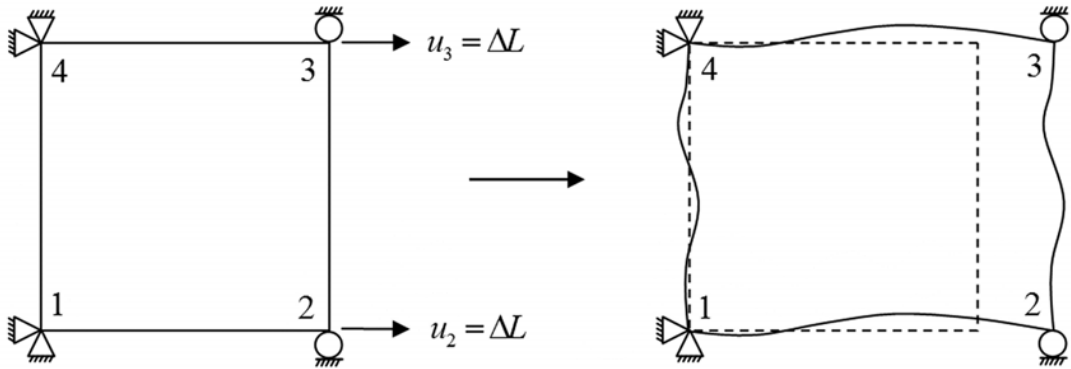

Figure 6: Boundary conditions imposed on the RVE cell mesh, required to produce a constant-width applied strain in the $x$-direction, both (a) before and (b) during deformation. The corner nodes have prescribed displacements, and the displacements of all other boundary nodes are related to those of the equivalent node on the opposite edge.

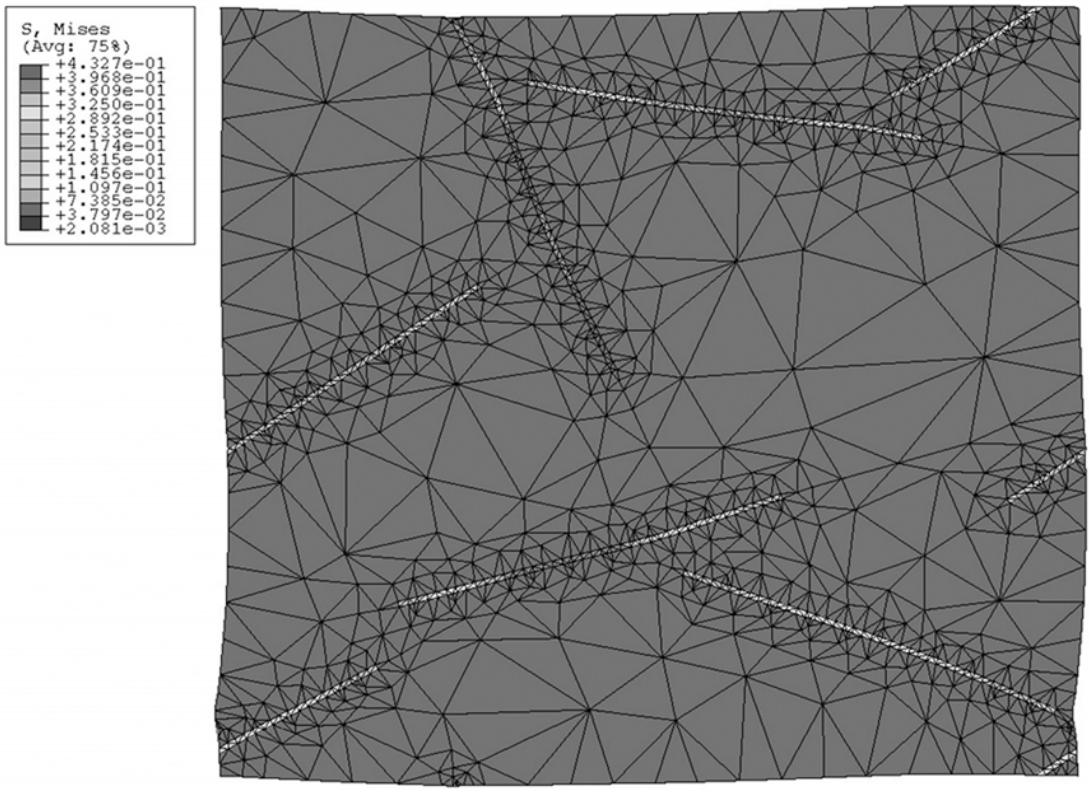

(a)

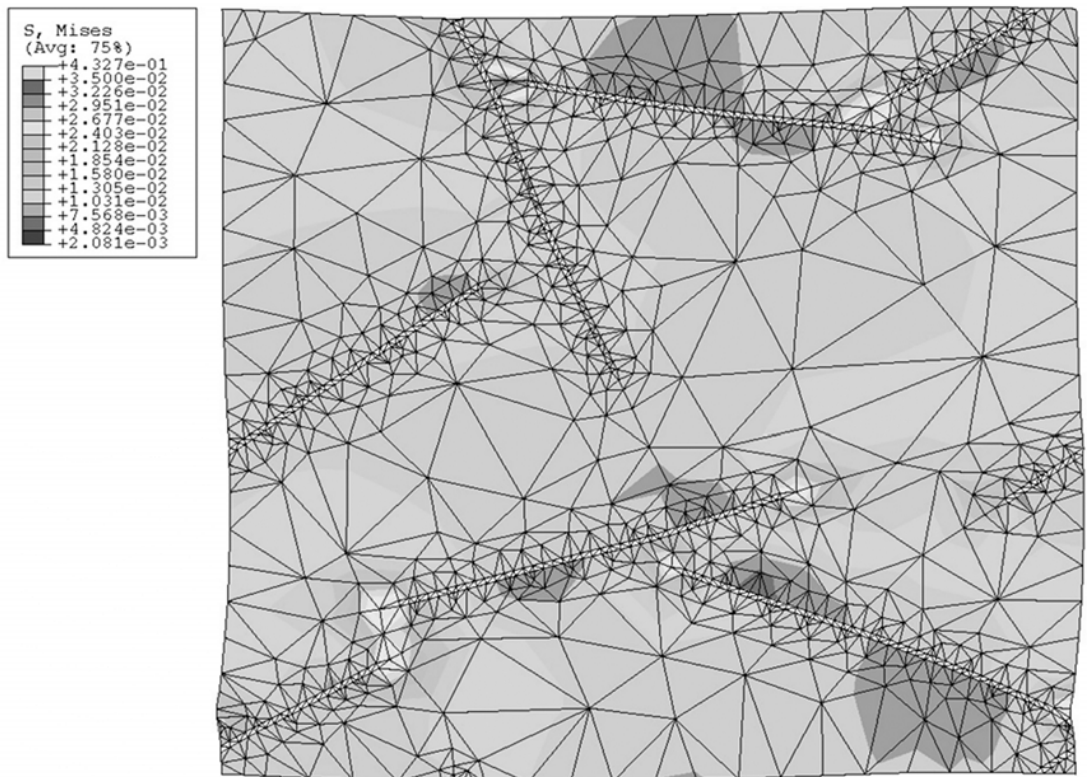

(b)

Figure 7: (a) The Mises stress resulting from a 1% constant-width *x*-stretch. The stress is taken up by the platelets. (b) The same cell geometry but with the scale altered so that the variation of the Mises stress in the polymer is visible.

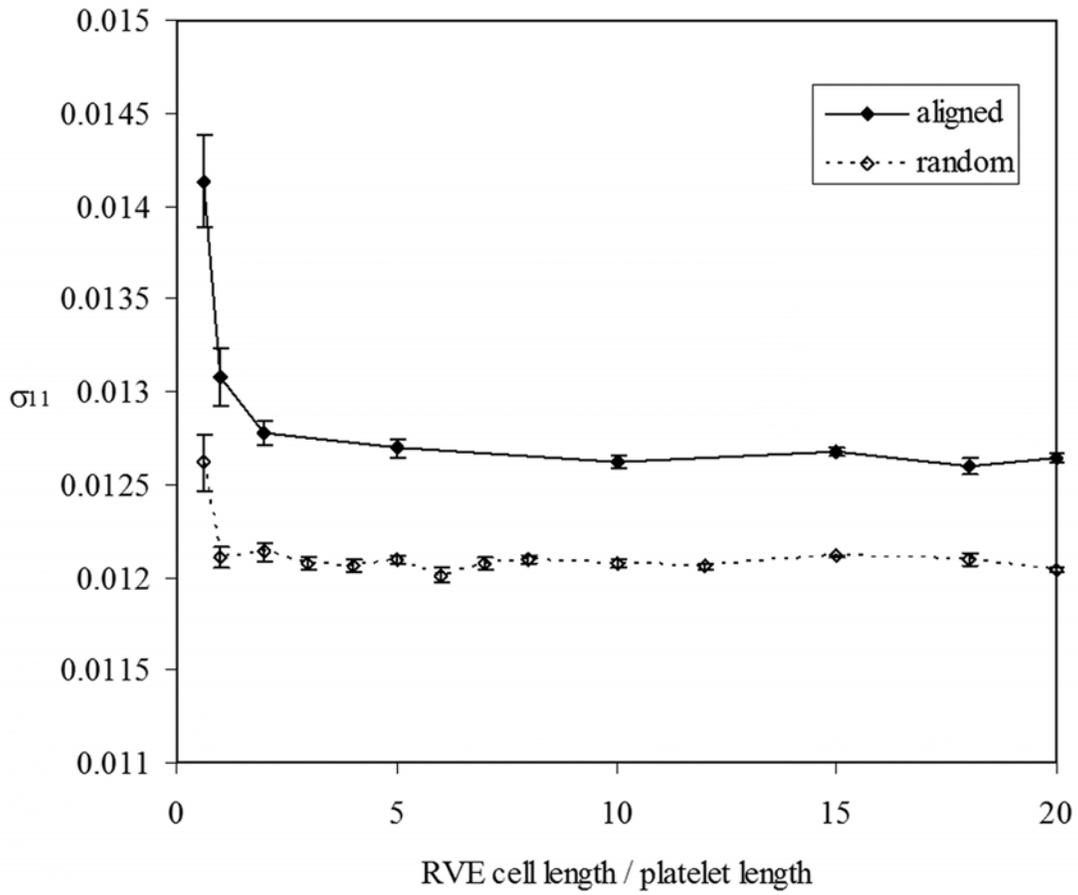

Figure 8: Convergence of the effective stress with increasing RVE cell size, for straight platelets of aspect ratio 100:1 and an area filling fraction of 4%. Each point is averaged over an ensemble of 25-250 realisations. Typically, we took the typical RVE cell size to be 5 times the platelet length.

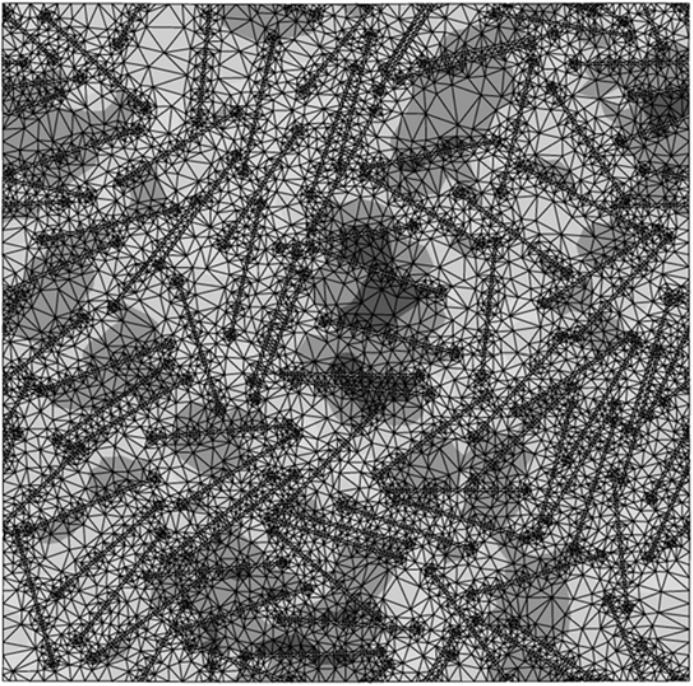

Figure 9: A typical RVE realisation cell for randomly orientated and randomly distributed single layer straight platelets, on application of a 1% constant-width *x*-strain.

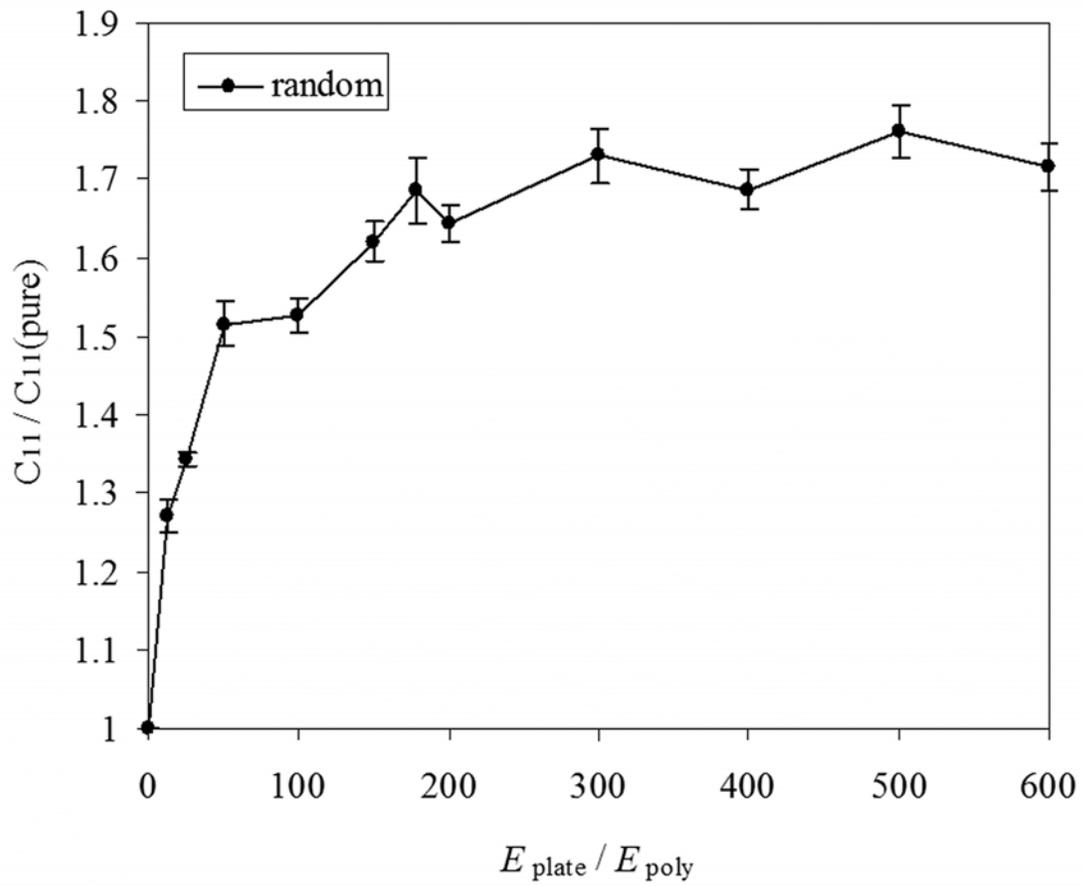

Figure 10: Variation of the stiffness enhancement $C_{11}/C_{11}(\text{pure})$ with the platelet stiffness for randomly orientated straight platelets. We expect the platelet stiffness to be at least $E_{plate}/E_{poly} \approx 200$ in the real material.

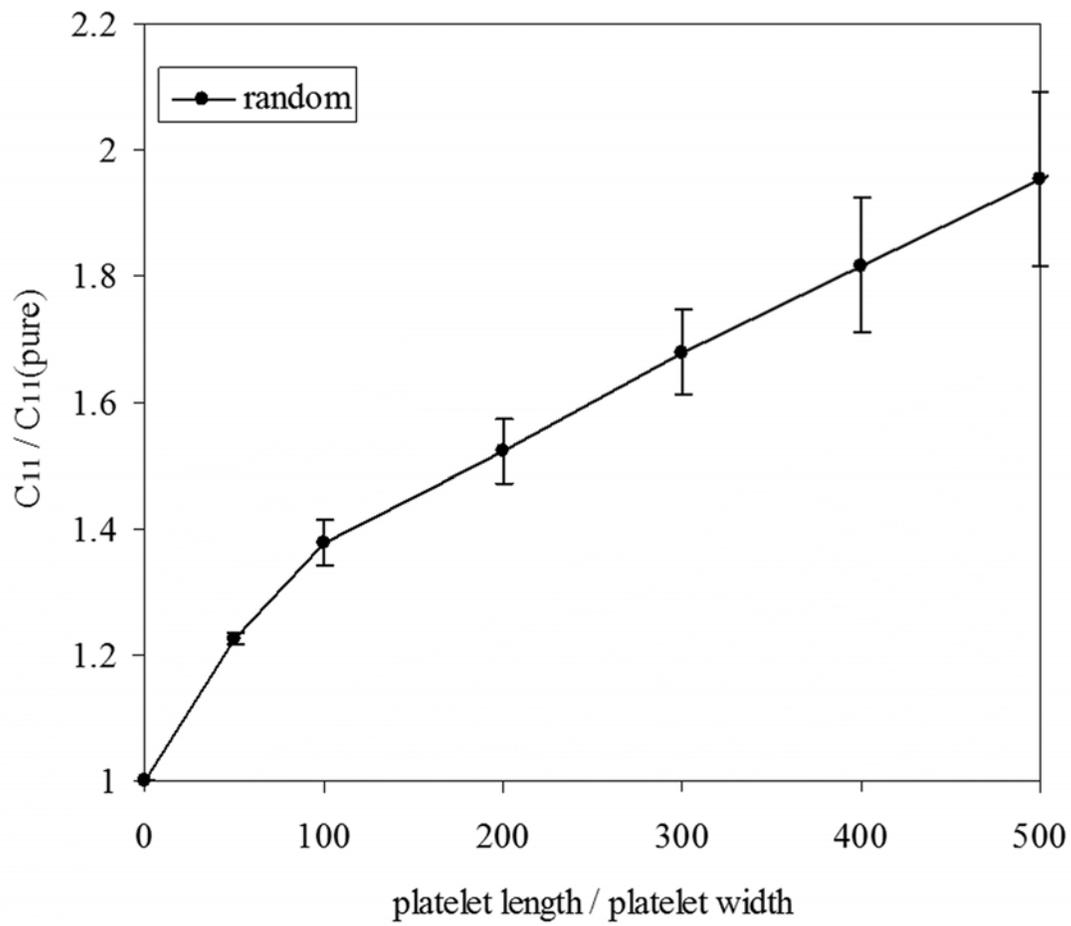

Figure 11: Variation of stiffness enhancement with changing aspect ratio, for randomly orientated straight platelets, with a fixed area filling fraction of 4%.

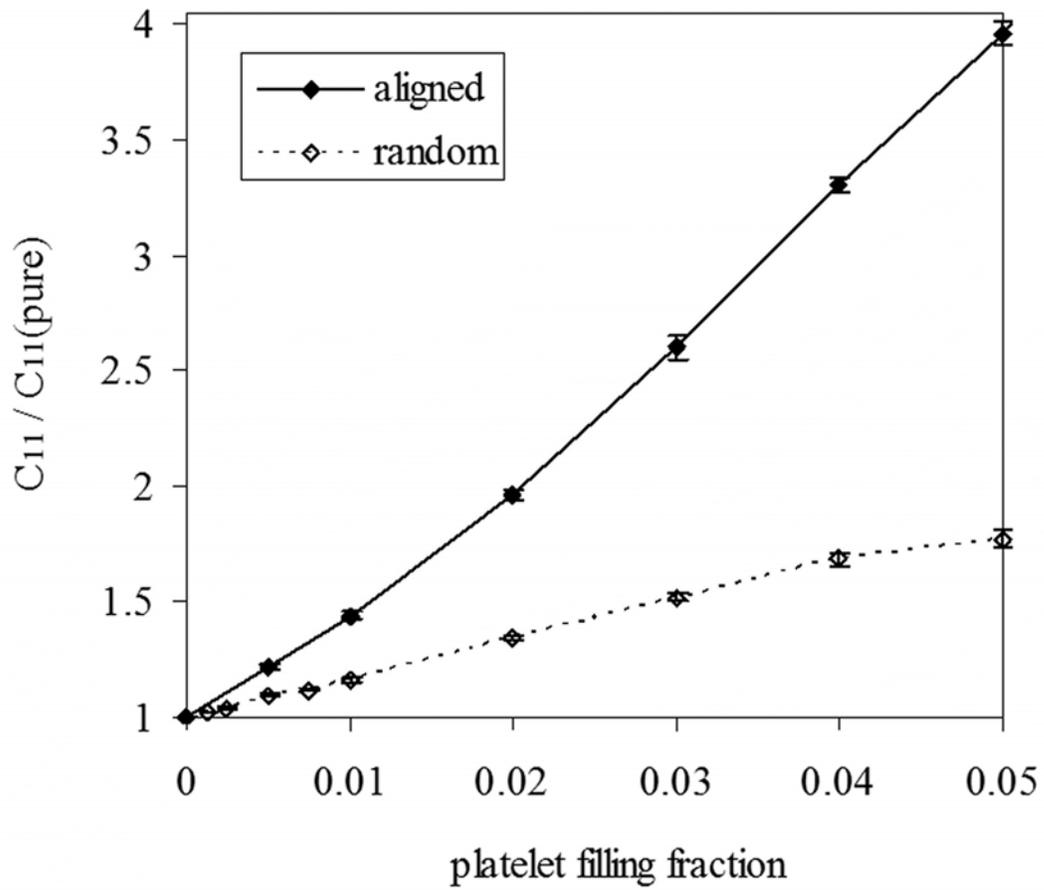

Figure 12: Enhancement in stiffness versus the area platelet filling fraction, for both randomly aligned platelets, and platelets aligned with the *x*-direction. The platelets were straight and had a fixed aspect ratio (length/width) of 100:1.

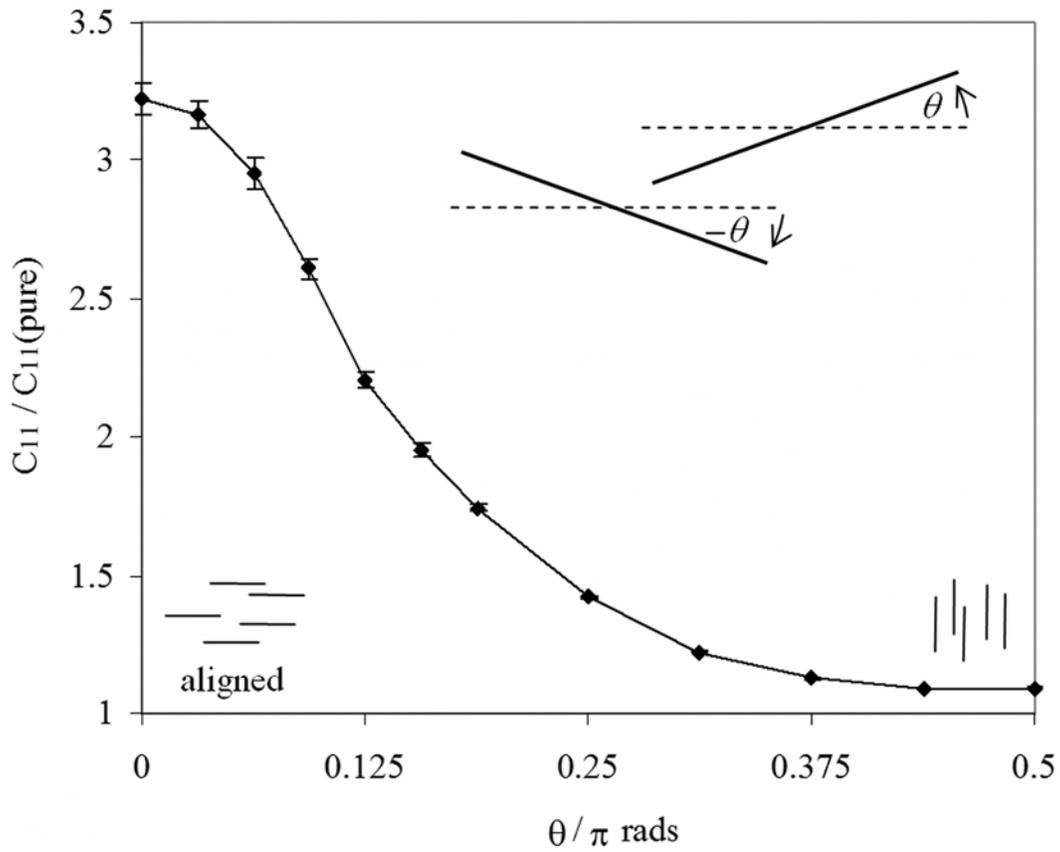

Figure 13: Stiffness enhancement for RVE geometries in which all platelets have a fixed tilt angle of $\pm\theta$ to the *x*-direction, for straight platelets with a fixed aspect ratio of 100:1 and filling fraction of 4%. On the left of the plot ($\theta = 0$) the platelets are aligned perfectly with the stretch-direction, and on the right ($\theta = 90°$) they are all aligned perpendicular to it.

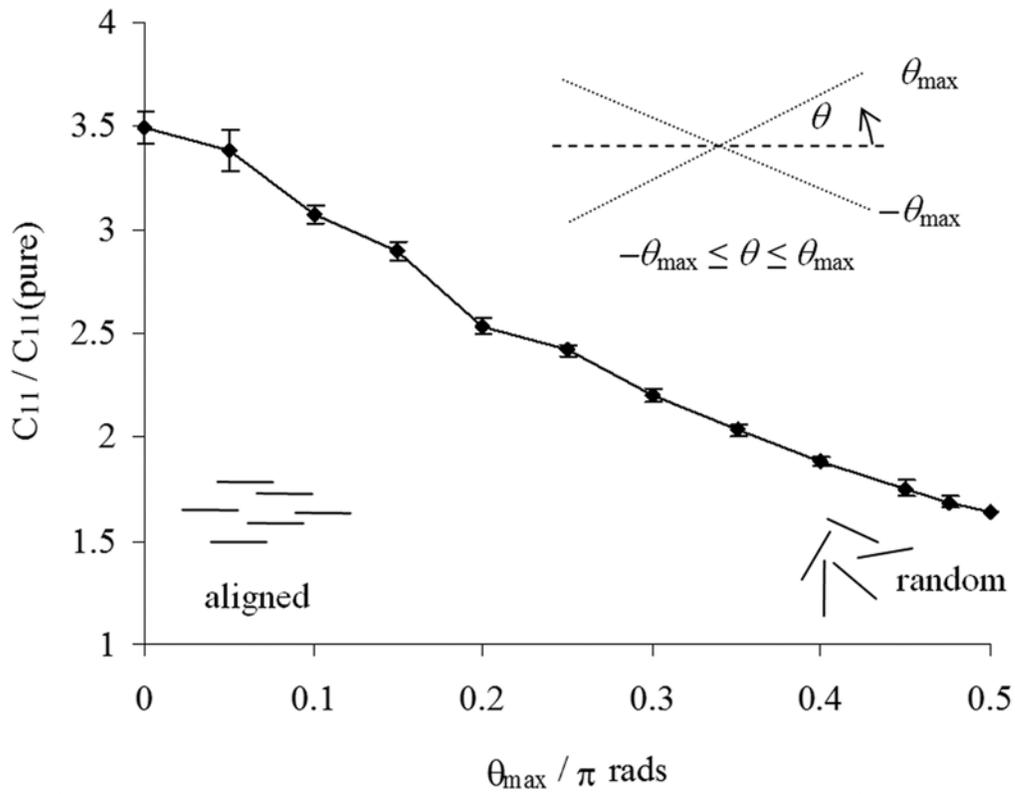

Figure 14: Enhancement for RVE geometries which have randomly-orientated straight platelets up to a maximum tilt angle of $\pm\theta$. On the left of the plot ($\theta_{max} = 0$) the platelets are all aligned with the $x$-direction, and on the right ($\theta_{max} = 90°$) the alignment of the platelets is completely random.

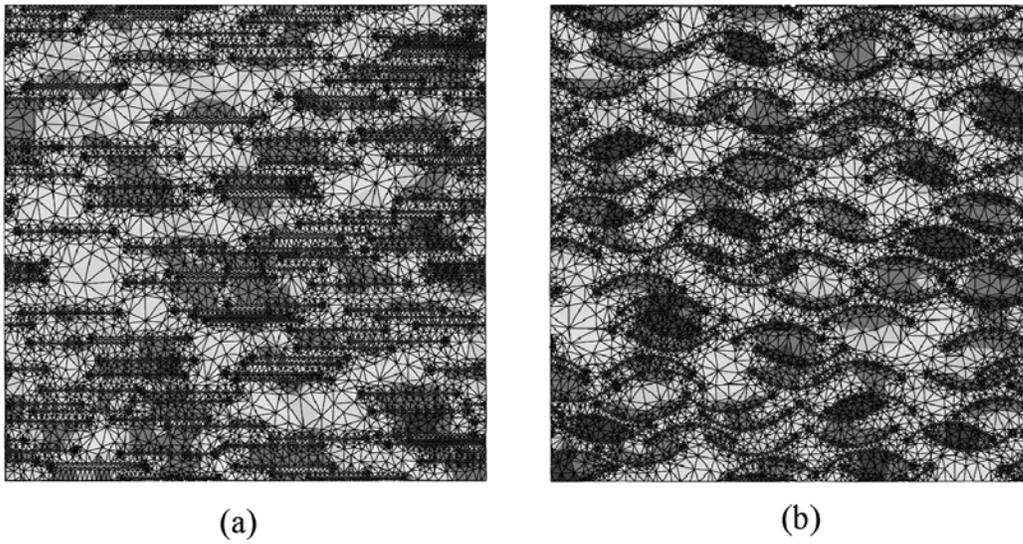

Figure 15: Typical RVE geometries for (a) straight platelets all aligned with the *x*-direction, and (b) curved platelets all aligned with the *x*-direction, for the same platelet filling fraction and aspect ratio.

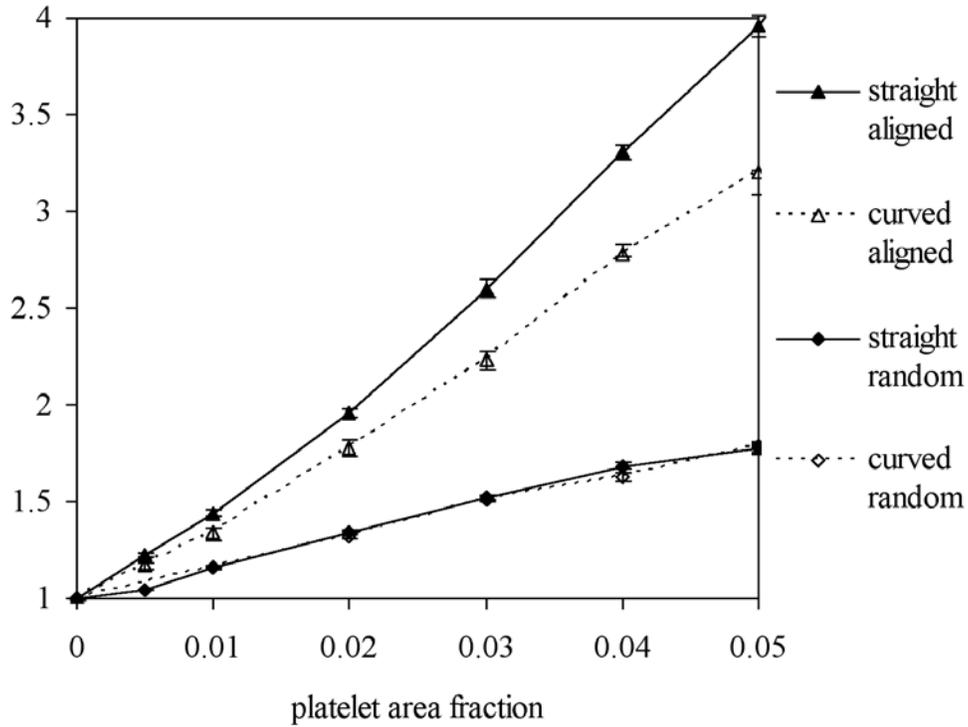

Figure 16: Stiffness enhancement for 4 different cases involving either all straight or all curved platelets that are either randomly orientated or aligned with the *x*-direction, at the same filling fraction of 4%, and aspect ratio of 100:1.

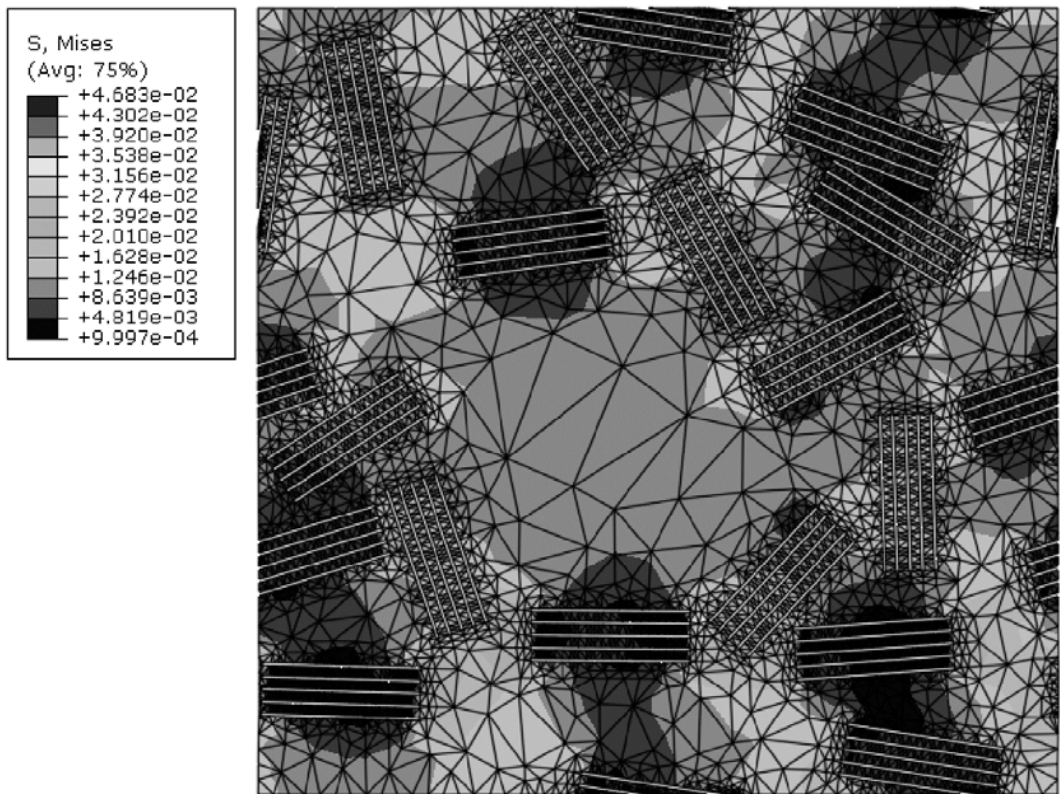

Figure 17: A typical RVE cell geometry containing randomly orientated and randomly distributed 5-layered particle-stacks, representing partial exfoliation.

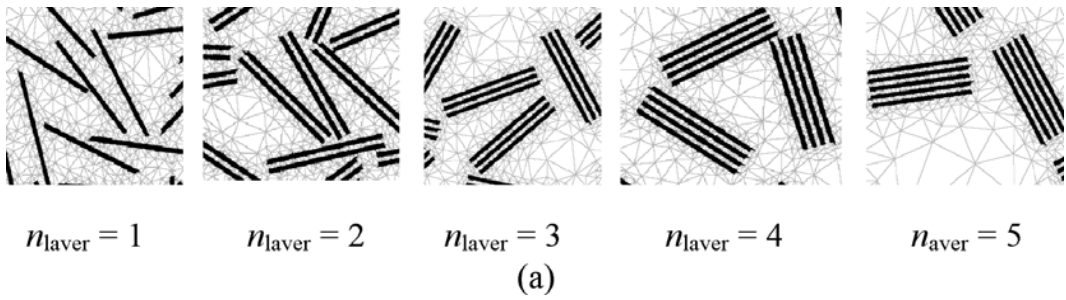

$n_{layer} = 1$    $n_{layer} = 2$    $n_{layer} = 3$    $n_{layer} = 4$    $n_{layer} = 5$

(a)

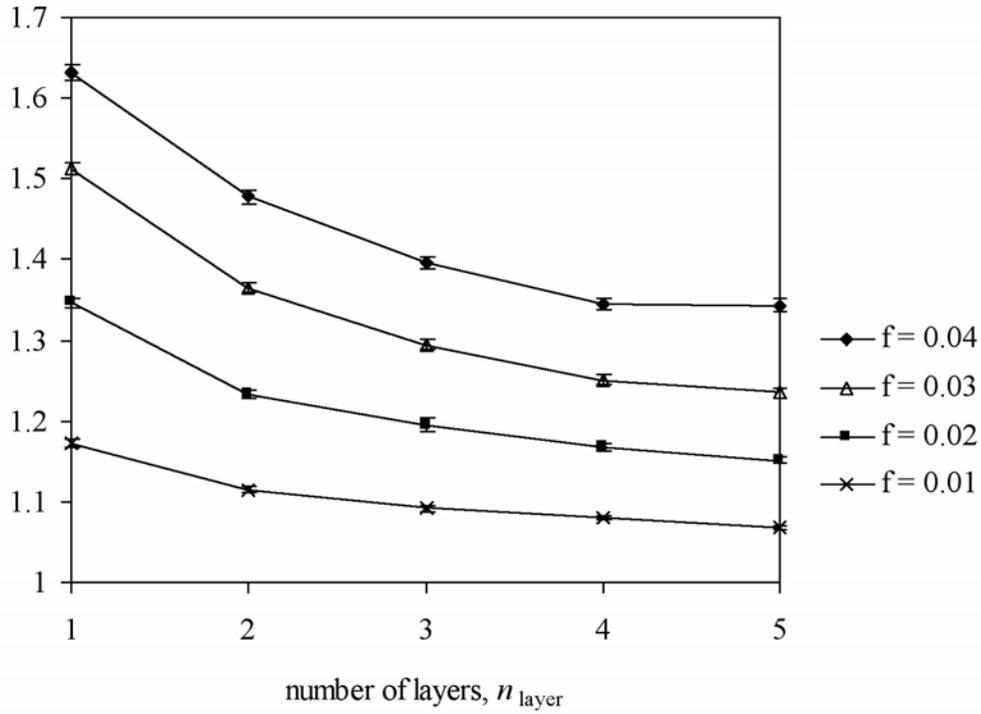

Figure 18: (a) Small sections of RVE cell involving stacks of 1,2,3,4, and 5 layers. (b) Stiffness enhancement versus the number of layers in the stack, for various area filling fractions $f$. The multilayer stacks are randomly aligned and randomly distributed.

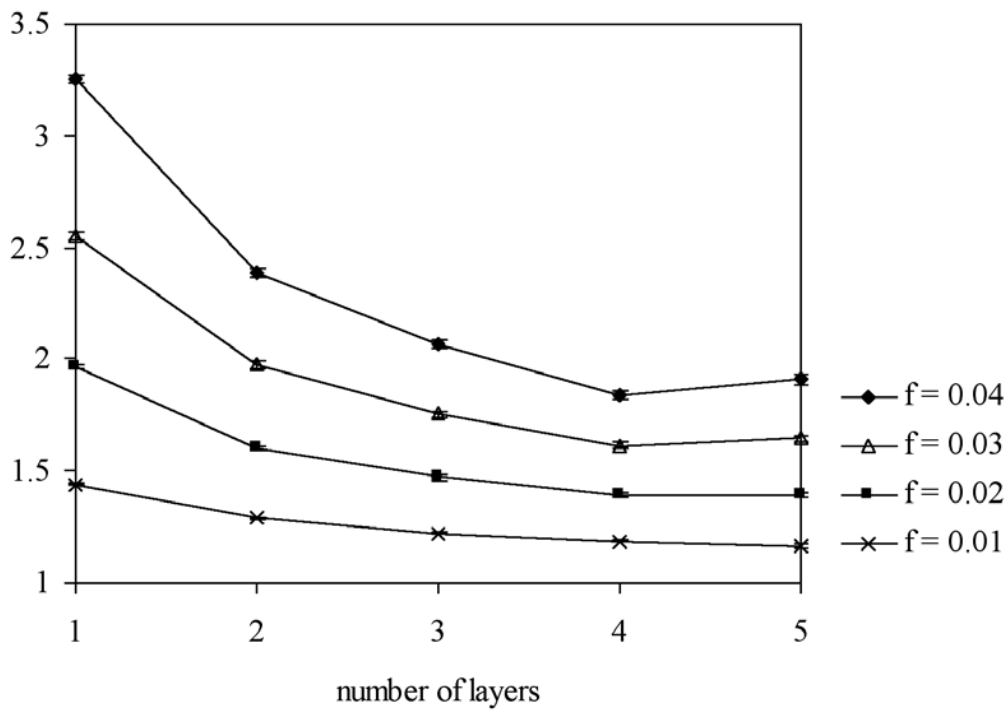

Figure 19: The variation in stiffness enhancement with the number if layers in the stack, at various area filling fractions, where all the particles are aligned with the *x*-direction.